\documentclass[usenatbib]{mn2e}
\usepackage{color}
\usepackage[tight]{subfigure}
\usepackage{longtable}
\usepackage{amssymb}
\usepackage{lscape}
\usepackage{graphicx}
\usepackage{natbib}
\usepackage{enumitem}
\usepackage{url}

\def \simless {\mathbin{\lower 3pt\hbox{$\rlap{\raise 4pt
              \hbox{$\char'074$}}\mathchar"7218$}}}
\def \simgreat {\mathbin{\lower 3pt\hbox{$\rlap{\raise 4pt
              \hbox{$\char'076$}}\mathchar"7218$}}}
\def\ie{{\it i.e.,}}
\def\cf{{\it c.f.,}}
\def\eg{{\it e.g.,}}

\title[Stellar populations of BCGs in cool-core clusters]{The regulation of star formation in cool-core clusters: imprints on the stellar populations of brightest cluster galaxies}
\author[Loubser et al.]{S. I. Loubser$^{1}$\thanks{E-mail:Ilani.Loubser@nwu.ac.za (SIL)}, A. Babul$^{2}$, H. Hoekstra$^{3}$, A. Mahdavi$^{4}$, M. Donahue$^{5}$,
\newauthor{C. Bildfell$^{2}$ and G. M. Voit$^{5}$}\\
$^{1}$Centre for Space Research, North-West University, Potchefstroom 2520, South Africa\\
$^{2}$Department of Physics and Astronomy, University of Victoria, Victoria, BC, V8W 2Y2, Canada\\
$^{3}$Leiden Observatory, Leiden University, PO Box 9513, 2300 RA, Leiden, The Netherlands\\
$^{4}$Department of Physics and Astronomy, San Francisco State University, San Francisco, CA 94131, USA\\
$^{5}$Michigan State University, Physics $\&$ Astronomy Dept., East Lansing, MI 48824-2320, USA}

\begin{document}

\date{Accepted 2015 November 24.  Received 2015 November 20; in original form 2015 August 23}

\pagerange{\pageref{firstpage}--\pageref{lastpage}} \pubyear{2015}

\maketitle

\label{firstpage}

\begin{abstract}
A fraction of brightest cluster galaxies (BCGs) Êshows Êbright  emission in the UV Êand the blue part of the optical spectrum, which has been interpreted as evidence of recent star formation.Ê Most of these results are based on the analysis of broadband photometric data. ÊHere, we study the optical spectra of a sample of 19 BCGs hosted by X-ray luminous galaxy clusters at $0.15<z<0.3$, a subset from the Canadian Cluster Comparison Project (CCCP) sample. We identify plausible star formation histories of the galaxies by fitting Simple Stellar Populations (SSPs) as well as composite populations, consisting of a young stellar component superimposed on an intermediate/old stellar component, to accurately constrain their star formation histories.  We detect prominent young $(\sim 200\;\rmn{Myr})$ stellar populations in 4 of the 19 galaxies.   Of the four, the BCG in Abell 1835 shows remarkable A-type stellar features indicating a relatively large population of young stars, which is extremely unusual even amongst star forming BCGs.   We constrain the mass contribution of these young components to the total stellar mass to be typically between 1\% to 3\%, but rising to 7\% in Abell 1835.   We find that the four of the BCGs with strong evidence for recent star formation (and only these four galaxies) are found within a projected distance of $5\;\rmn{kpc}$  of their host cluster's X-ray peak, {\it and} the diffuse, X-ray gas surrounding the BCGs exhibit a ratio of the radiative cooling-to-free-fall time ($t_{\rm c}/t_{\rm ff}$) of $\leq 10$. These are also some of the clusters with the lowest central entropy.   Our results are consistent with the predictions of the precipitation-driven  star formation and AGN feedback model, in which the radiatively cooling diffuse gas is subject to local thermal instabilities once the instability parameter $t_{\rm c}/t_{\rm ff} $ falls below $\sim 10$, leading to the condensation and precipitation of cold gas. The number of galaxies in our sample where the host cluster satisfies all the criteria for recent and ongoing star formation is small, but their stellar populations suggest a timescale for star formation to restart of the order of $\sim$ 200 Myrs.

\vskip+10pt

\end{abstract}


\section{Introduction}
\label{Sec1}


A fraction of the brightest cluster galaxies (BCGs) exhibit evolutionary histories that are in stark contrast with the conventional expectation that giant elliptical galaxies in clusters of galaxies are all quiescent, passively evolving, ``red and dead" systems. BCGs, particularly those residing at the centres of cool-core clusters \citep{Hu1985, McCarthy2004, McCarthy2008, Cavagnolo2008} exhibit ``blue cores" \citep{Bildfell2008} corresponding to the presence of a small fraction ($\sim$ 1 per cent) of very young stars \citep{Sarazin1983, Cardiel1995, Cardiel1998, Pipino2009}. These BCGs also reveal activity in their radio emission \citep{Burns1990, Best2007}, or have optical emission-line nebulae \citep{Crawford1999, Edwards2007, Loubser2013}, excess UV light \citep{McNamara1989, Rafferty2008, Donahue2010}, far-infrared emission from warm dust \citep{Quillen2008}, and molecular gas \citep{Edge2002, Rawle2012}. Nevertheless, this forms far from a complete picture as BCGs in the nearby Universe exhibit diverse morphologies, differences in their stellar populations, and therefore star formation histories \citep{Loubser2009, Donahue2015}, despite the fact that most are located in a similar, privileged environment in the very centres of X-ray luminous clusters.  Short radiative cooling times for the diffuse gas in the cluster cores, and proximity of the BCG to the X-ray peak seem to be necessary \citep{Bildfell2008}, but not sufficient criteria for star formation in BCGs.   

Nearly 50\% of galaxy clusters in X-ray selected samples have central gas cooling times  $\simless\ 1$ Gyr \citep{Hudson2010} and if unchecked, radiative losses would engender a flow of cooled gas into the BCG resulting in star formation rates at least an order of magnitude higher (100 -- 1000 $\rmn{M_{\odot}/yr}$) than observed \citep{Fabian1991, Edge2001, ODea2008}.    The BCGs often host radio-loud AGN \citep{Sun2009}, and theoretical estimates have long suggested that the mechanical heating of the intracluster gas by these AGNs plays a key role in counteracting radiative cooling \citep{Binney1995, Ciotti2001, Soker2001, Babul2002}.   With imaging X-ray observations showing clear signs of interactions between the AGN outflows and the hot diffuse gas surrounding the BCGs  (see \citealt{McNamara2007, McNamara2012} and references therein), there is now a broad consensus that in order to make sense of the full range of BCG properties, it is crucial to understand the role of cooling, feedback and gas accretion in cluster cores.   However, the details of how this coupling operates and of the precise mechanism(s) linking the different phenomena remain a puzzle.

One possibility is that AGN feedback is directly fueled by the flow of the hot diffuse gas onto the central black hole via the hot-mode (Bondi) accretion.   The main argument in favour of this mode is that the AGN activity is directly tied to the state of the diffuse gas because the Bondi accretion rate depends on the gas density  and  temperature  in the vicinity of the black hole \citep{Allen2006}.   For this reason, this is the fueling mode most commonly invoked in cosmological simulations of galaxies, galaxy groups and clusters of galaxies (\eg~   \citealt{DiMatteo2005, Fabjan2010, Dubois2011, Bellovary2013, LeBrun2014}).  However,  a detailed analysis of AGN activity in galaxy clusters \citep{Raffert2006, McNamara2011} indicates that the hot-mode accretion is much too small (by orders of magnitude) to account for the observed outbursts (but see \citealt{Fujita2014}).  \citet{Hardcastle2007} also come to the same conclusion in their study of powerful radio jets.

More recently, an alternative ``cold mode accretion model'' \citep{Pizzolato2005, McCourt2012, Sharma2012, Gaspari2013} has emerged as a useful framework for interpreting various observed properties of elliptical galaxies as well as BCGs in the centres of clusters \citep{Gaspari2012, Li2014a, Li2014b,Prasad2015, Voit2014, Voit2015}.  In this model, the diffuse X-ray emitting gas in the cluster centre, which exists in approximate global balance between radiative cooling and AGN feedback, is expected to become highly susceptible to local thermal instabilities once the ratio of the radiative cooling-to-free-fall time  ($t_{\rm c}/t_{\rm ff}$) drops below a threshold value of $\sim 10$ \citep[\cf][]{Singh2015}, resulting in  cold dense clouds condensing out of the diffuse gas and the emergence of a multiphase circumgalactic medium.   Once the cold clouds begin to form, recent simulation studies show that these typically fall freely towards the cluster centre and, in the process, give rise to tens of kpc long cold filaments threading the BCG and its outskirts as well as to an accumulation of cold gas in the centres of the BCGs \citep{Gaspari2012, Li2014a, Prasad2015}.  The latter, in turn, is expected to fuel AGN outbursts as well as star formation activity in the galaxies. 

In a recent study of a large sample of {\it ACCEPT} clusters, \citet{Voit2014} assessed the relationship between the existence and the prominence of the extended emission-line nebulae, as evidenced by the observed H$\alpha$ emission, in the cluster cores and the thermodynamic state of the diffuse, X-ray emitting gas cocooning the BCG.   They found that not only were all the clusters with min($t_{\rm c}/t_{\rm ff}) < 20$ detected in H$\alpha$, but that this thermal instability criterion also correlated with the measured H$\alpha$ luminosity in the sense that clusters with small values of  min($t_{\rm c}/t_{\rm ff})$ are the most luminous.   This is consistent with the model prediction that the more susceptible the gas is to thermal instabilities, the greater the total amount of multiphase gas in the cluster cores.

In this paper, we investigate the other model prediction, namely that star formation activity in the BCGs is also intimately linked to the stability of the X-ray emitting gas in the cluster cores.   Specifically,  we use the optical spectra of BCGs located in X-ray luminous clusters at 0.15 $< \rmn{z} <$ 0.3 in combination with the stellar population synthesis models to establish their diverse star formation histories (SFHs). We explore whether there is a clear relationship between these histories and the environmental condition within the cluster cores.  In section 2 below, we describe the optical spectroscopy and data reduction.    In section 3, we describe the stellar population analysis, and we discuss the results in section 4.   We compare the results with UV data and theoretical hypotheses in section 5, and we summarise the main results in section 6.    Throughout this paper we adopt a $\Lambda$CDM cosmology with $\Omega_{m} = 0.3$, $\Omega_{\Lambda} = 0.7$, and $\rmn{H_{0} = 70\ km\ s^{-1} Mpc^{-1}}$. 
 

\section{Optical spectroscopy and data reduction}
\label{reduction}

The Canadian Cluster Comparison Project (CCCP) consists of a sample of 50 clusters studied in \citet{Mahdavi2013} (see also \citealt{Hoekstra2012, Hoekstra2015}). We discard potential double clusters (Abell 115N/S, Abell 223 N/S, Abell 1758E/W), as well as clusters without a clearly dominant BCG (Abell 959, MS 1231+15a/b, Abell 1234, Abell 1246; \cf \ \citealt{Bildfell2008} for details). We target clusters in the redshift range $0.15 < z < 0.3$ whose BCGs reside within a projected distance of 75 kpc of their host cluster's X-ray peak. Of the 24 clusters satisfying this criteria, we observed 19 using the GMOS detector in long-slit mode over a period of three semesters from 2007B to 2008B on the Gemini North and South observatories. The clusters that we did not observe in the allocated time, or for which the acquired data are not of sufficient quality for the analysis discussed here, were Abell 2204, Abell 2218, Abell 2219, Abell 521, Abell 697 (see Table \ref{bcgs}). These are all listed as red cored galaxies in \citep{Bildfell2008}, except Abell 2204 which is a known blue core BCG. As discussed in section 5, we do not expect these five systems to change our conclusions.

We use the B600 grating with a central wavelength of 460 $\rmn{nm}$ and a slit width of 0.75 $\rmn{arcsec}$. The slit is aligned with the major axis of the BCG.  We use a 2 $\times$ 2 on-instrument pixel binning which, along with the instrument configuration described above, yields a spectral pixel scale of 0.1 $\rmn{nm/pixel}$ and a spatial pixel scale of 0.14 $\rmn{arcsec/pixel}$. For each cluster we take four separate 1800 $\rmn{s}$ observations, using small spatial and spectral dithers to mitigate problems introduced by detector defects and chipgaps. There are some exceptions however, as due to scheduling and weather constraints one cluster was observed for less time (2 $\times$ 1800 $\rmn{s}$), while two others were observed for more time (7 $\times$ 1800 $\rmn{s}$ and 8 $\times$ 1800 $\rmn{s}$). Further information on our observations is given in Table \ref{bcgs}.

\begin{table*}
\begin{tabular}{l c c c c r r r r r}
\hline
Name & $z$ & $\alpha_{J2000}$& $\delta_{J2000}$ & Exp. time & $R_{off}$ & $\%$ of $R_{e}$ & $t_{c, 0}$ & $K_{0}$ & $t_{c}/t_{ff}$\\
 & & & & $\rmn{(ks)}$ & $\rmn{(kpc)}$ & & $\rmn{(Gyr)}$ & $\rmn{(keV cm^{2})}$ & \\
\hline
Abell 68 & 0.26 & 00:37:06.85 & +09:09:24.51 & 4$\times$1.8 & 14.0$\pm$2.4 & 0.18 & 3.57$\pm$0.21 &214.2$\pm$35.3 &57.0$\pm$28.0\\
Abell 209 & 0.21 & 01:31:52.54 & -- 13:36:40.00 & 4$\times$1.8 & 14.4$\pm$2.5 & 0.17 & 3.54$\pm$0.18 & 160.8$\pm$19.8&90.6$\pm$33.8\\
Abell 267 & 0.23 & 01:52:41.95 & +01:00:25.89 & 4$\times$1.8 & 65.2$\pm$5.0 & 0.20 & 4.15$\pm$0.37 &152.7$\pm$34.1&109.0$\pm$55.7\\
Abell 383 & 0.19 & 02:48:03.38 & -- 03:31:44.93 & 7$\times$1.8 & 0.6$\pm$2.0 & 0.56 & 0.41$\pm$0.02 & 21.3$\pm$1.0&8.5$\pm$0.3\\
Abell 568 & 0.17 & 07:32:20.31 & +31:38:01.06 & 8$\times$1.8 & 10.6$\pm$2.1 & 0.26 & 2.86$\pm$0.31 &140.1$\pm$23.2&32.7$\pm$8.5\\
Abell 611 & 0.29 & 08:00:56.83 & +36:03:23.79 & 4$\times$1.8 & 3.4$\pm$2.0 & 0.44 & 1.28$\pm$0.11 &57.0$\pm$10.0&30.9$\pm$6.5\\
Abell 963 & 0.21 & 10:17:03.63 & +39:02:49.67 & 4$\times$1.8 & 5.5$\pm$2.0 & 0.18 & 1.32$\pm$0.06 &63.1$\pm$4.7&28.6$\pm$2.3\\
Abell 1689 & 0.18 & 13:11:29.52 & -- 01:20:27.86 & 4$\times$1.8 & 3.7$\pm$2.0 & 0.32&1.19$\pm$0.04 &72.5$\pm$3.4&18.2$\pm$0.6\\
Abell 1763 & 0.22 & 13:35:20.12 & +41:00:04.30 & 4$\times$1.8 & 7.8$\pm$2.0 & 0.60 & 10.61$\pm$1.05 &419.5$\pm$54.0& --- \\
Abell 1835$^{\star}$ & 0.25 & 14:01:02.10 & +02:52:42.69 & 4$\times$1.8 & 4.5$\pm$2.0 & 0.28 & 0.29$\pm$0.01 &19.7$\pm$0.4&3.4$\pm$0.1\\
Abell 1942$^{\star}$ & 0.22 & 14:38:21.88 & +03:40:13.34 & 4$\times$1.8 & 5.0$\pm$2.0 & 0.35 & 6.26$\pm$0.98 &230.6$\pm$79.3&124.3$\pm$55.7\\
Abell 2104 & 0.15 & 15:40:07.94 & -- 03:18:16.25 & 4$\times$1.8 & 6.4$\pm$2.1 & 0.27 & 5.52$\pm$0.68 &201.7$\pm$37.6&83.7$\pm$50.3\\
Abell 2259 & 0.16 & 17:20:09.66 & +27:40:08.29 & 2$\times$1.8 & 72.5$\pm$8.8 & 0.34 & 3.71$\pm$0.40 &134.7$\pm$33.6&82.2$\pm$45.4\\
Abell 2261 & 0.22 & 17:22:27.23 & +32:07:57.72 & 4$\times$1.8 & 0.3$\pm$2.0 & 0.42 & 1.14$\pm$0.11 &60.0$\pm$8.7&27.8$\pm$5.1\\
Abell 2390$^{\star}$ & 0.23 & 21:53:36.84 & +17:41:44.10 & 4$\times$1.8 & 2.4$\pm$2.0 & 0.83 & 0.58$\pm$0.01 &31.6$\pm$1.0&9.8$\pm$0.2\\
Abell 2537 & 0.30 & 23:08:22.22 & -- 02:11:31.74 & 4$\times$1.8 & 15.1$\pm$2.3 & 0.33 & 2.05$\pm$0.25 & 91.8$\pm$20.0&32.8$\pm$14.2\\
MS 0440+02 & 0.19 & 04:43:09.92 & +02:10:19.33 & 4$\times$1.8 & 0.9$\pm$2.0 & 0.15 & 0.73$\pm$0.07 &30.1$\pm$6.8&11.7$\pm$4.0\\
MS 0906+11 & 0.17 & 09:09:12.76 & +10:58:29.12 & 4$\times$1.8 & 2.8$\pm$2.0 & 0.28 & 2.88$\pm$0.45 &148.9$\pm$22.5&46.0$\pm$20.3\\
MS 1455+22 & 0.26 & 14:57:15.12 & +22:20:34.48 & 4$\times$1.8 & 3.6$\pm$2.0 & 0.63 & 0.40$\pm$0.01 & 23.6$\pm$0.6&6.3$\pm$0.2\\
\hline
Abell 521 & 0.25 &  04:54:06.86 & -- 10:13:24.50 & -- & 31.3$\pm$5.0 & -- & 2.43$\pm$0.37 &75.6$\pm$17.7&	112.5$\pm$28.3\\
Abell 697 & 0.28 &  08:42:57.54 & +36:21:59.90 & -- & 13.9$\pm$2.6 & -- & 5.88$\pm$1.12 & 240.0$\pm$48.8 & 206.4$\pm$39.1\\
Abell 2204 & 0.15 & 16:32:46.95 & +05:34:33.10 &  -- & 0.8$\pm$2.0 & -- & 0.25$\pm$0.00 & 17.3$\pm$0.3 & 8.4$\pm$1.2\\
Abell 2218 & 0.18 &  16:35:49.22 & +66:12:44.80 & -- & 56.4$\pm$4.3 &  -- &8.45$\pm$1.38   & 317.9$\pm$47.1 & 150.7$\pm$38.5\\
Abell 2219 & 0.23 &  16:40:19.83 & +46:42:41.50 & -- & 6.8$\pm$2.1 & -- & 5.84$\pm$0.77 & 243.2$\pm$30.6 & 89.6$\pm$19.4\\
\hline
\end{tabular}
\caption{BCGs observed and analysed for this study. Objects marked with $\star$ overlap with the analysis in \citet{Pipino2009}. $R_{off}$ is the projected distance between the BCG and the X-ray peak \citep{Bildfell2008}, and we list the fraction of the half-light radius, $R_{e}$, reached with the extracted 15 $\rmn{kpc}$ apertures. The entropy $K_{0}$ and the cooling time $t_{c, 0}$ at a radius of 20 $\rmn{kpc}$ (hereafter referred to as ``central'') are from high-resolution \textit{Chandra, XMM-Newton} X-ray data analysis \citep{Mahdavi2013}. We describe the derivation of the  $t_{c}/t_{ff}$ ratios in section \ref{clusters}. The five galaxies below the line were not observed and/or analysed (as described in section \ref{reduction}), but we do not expect these to affect our conclusions as described in section 5.}
\label{bcgs}
\end{table*}

We performed the primary data reduction steps for spectral observations with the standard \textsc{iraf} software package for reducing GMOS data provided by Gemini. The basic reductions included bias-subtraction, flat-fielding and illumination correction. We use the Cu-Ar arc spectra to generate a set of dispersion functions sampled uniformly at several positions along the slit (ie. in the y-direction). For individual arc spectra, the residuals for known locations of Cu-Ar lines compared to the line locations, as predicted by the dispersion solutions, show a median RMS of 0.14 \AA{}. This translates to a worst-case systematic error of 11 $\rmn{km\ s^{-1}}$ for individual lines, but since our results are based on fitting the full wavelength range, for galaxies that have velocity dispersions of 300 $\rmn{km\ s^{-1}}$, the contribution of this source of systematic error is negligible. The dispersion solutions are applied to the corresponding science frames and standard star observations to obtain a set of wavelength-calibrated data. 

The next step in the reduction procedure is sky subtraction. The projected angular scale of the slit on the detector extends for a total of 5.6 $\rmn{arcmin}$. Avoiding the edges of the detector and the central 1.1 $\rmn{arcmin}$ region containing the spectrum of the BCG, we sample the spectra from a 1.9 $\rmn{arcmin}$ region below the BCG and a 1.8 $\rmn{arcmin}$ region above the BCG and sum these together. The extracted spectra are then fit column-by-column (ie. at fixed wavelength) with a linear relation to obtain a background model that varies as a function of wavelength and position along the slit. The background model is then subtracted from the entire image and the process is repeated for all science frames and standard star observations. We note that due to our long exposures, the brightest sky line [OI]$\lambda$5577 is saturated in many of the images and as a result is not properly subtracted. To prevent this from influencing our results, this region of the spectrum is masked out in any spectrum fitting analysis. 

The final step in the reduction of the individual science frames is flux calibration by means of an observed spectrophotometric standard star to generate a sensitivity function. The sensitivity function is then applied to the corresponding science frame observations to transform the spectra in units of counts on the detector to a set of spectra in physical units of $\rmn{erg\ cm^{-2}\ s^{-1}\ \AA{}^{-1}}$. Additionally, we apply a second-order correction for atmospheric extinction at this stage, the specific wavelength dependence of which is supplied by the Gemini observatory via data tables in the Gemini \textsc{iraf} reduction package.

Using the individual exposures and the combined image we estimate the errors on a pixel-by-pixel basis. The error for the median-combined image at each pixel ($\sigma_{f}$) is calculated by taking the standard deviation of the pixel values. To prevent cosmic rays from biasing the errors, we use an iterative sigma-clipped estimator, rejecting values that lie outside $\pm$ 10$\sigma_{f}$ from the median. This creates an array of the same size and dimensions as the combined image. We fit the flux error ($\sigma_{f}$) as a function of wavelength on a row-by-row basis, using a linear relation. We use the expected values from the best fit solutions to build a $\sigma_{f}(\lambda, y$) for each combined image. The resulting 2-D error array is stored as an error image for use in the spectral fitting procedure. 


After taking into account the typical size of a blue stellar core \citep{Bildfell2008}, the putative size of the ionisation contribution from the AGN (if present), the size of the seeing disc and the typical distance between the core of the BCG and the X-ray peak, we extract two spatial regions from the 2-D spectra. The first bin is the central 5 $\rmn{kpc}$ on either side of the luminosity peak of the galaxy (a width of 10 $\rmn{kpc}$ in total), whilst the second bin extends from 5 $\rmn{kpc}$ to 15 $\rmn{kpc}$ (both sides of the galaxy combined). These inner and outer apertures contain data with sufficient S/N to accurately determine stellar population parameters. At the highest redshift in our sample ($z$ = 0.3) the inner aperture represents an angular size of 1.2 $\rmn{arcsec}$, which is larger than the size of FWHM of the worst-case PSF for our data.

We adjust the population models to the changing resolution (with wavelength) of the spectrograph. We used the spectrophotometric standard star observed with the same instrumental configuration to derive the line-spread function (LSF) at regular wavelength intervals. We compute the physical dispersion of the galaxies, after we account for the stellar library template dispersion and instrumental broadening. We present the spatially resolved velocity dispersion profiles of the BCGs, together with stellar dynamical modelling in conjunction with the X-ray and weak-lensing derived mass distributions of the host clusters in a subsequent article.

\section{Stellar populations}
\label{stellarpops}

The integrated (unresolved) spectra from passively evolving BCGs can be represented as Single Stellar Populations (SSPs, entities that consist entirely of stars born at the same time with the same metallicity). Modern stellar population analysis assumes that the stellar populations of a galaxy with a more complex SFH consist of a sum of SSPs. Stellar evolution theory is then used to calculate the integrated properties of these SSPs and predict a spectral energy distribution that can be compared to observations.  

We follow the methods of \citet{Koleva2009}, \citet{Loubser2014} and \citet{Groenewald2014} to analyse the stellar populations of BCGs. We use the University of Lyon Spectroscopic analysis Software (ULySS) \citep{Koleva2009} to obtain a non-linear $\chi^{2}$--minimization against SSP (or a linear combination of more than one SSP) models convolved with the internal kinematics of the galaxy. We use two sets of high-resolution stellar population models, those of \citet{Vazdekis2010} based on the MILES library \citep{Sanchez2006} and Pegase-HR \citep{Leborgne2004} based on the Elodie 3.2 library \citep{Prugniel2001}. Emission lines (if present) are masked, as even weak emission lines can fill-in important absorption features like the Balmer lines and will lead to overestimated ages. The rest of the spectrum is then used to determine the best fit to the stellar population models. We found consistent fits using both sets of models, and continue with the high-resolution Vazdekis/MILES models, using a Salpeter IMF \citep{Salpeter1955}, throughout the analysis. The errors on the stellar population parameters are derived by perturbing the spectrum with noise following a Gaussian distribution of width given by the error spectrum.

We start by identifying both the best-fitting spectrum resulting from an SSP (of age $t_{SSP}$) as well as the best-fitting composite stellar model comprising of a stellar population of mass fraction $f_{young}$ and age $t_{young} < 2$ Gyr superposed on an intermediate/old stellar population with $t_{old} > 2$ Gyr. A stellar spectrum with an age below 2 Gyr has prominent features that disappear in spectra of ages above 2 Gyr. Thus a spectrum with a young component (even if its mass fraction is small, its contribution to the luminosity can be significant) can not be properly fit with an SSP, or a combination of populations older than 2 Gyr. 

If both (SSP and composite model) give comparable $\chi^{2}$--values, we default to an SSP spectrum and interpret the results to indicate that (1) there is no significant young ($< 2$ Gyr) component present although  (2) if the $t_{SSP} < 10$ Gyr then the stellar population consists of both an old stellar population as well as intermediate aged stellar populations. Attempting to fit a composite model where all the components are older than 2 Gyr results in multiple, degenerate (with similar $\chi^{2}$ values) solutions due to the similarity in optical spectral features of older stellar populations. Given the errors in the observed spectra, it is more reliable to default to an SSP. An SSP fit is heavily luminosity-weighted, and therefore the presence of a small younger (luminous) component will skew the best SSP fit towards a younger age. For this reason, it is important to do a relative comparison between the parameters determined for the inner vs outer apertures, as well as a comparison of the stellar population properties among the galaxies in the sample, and the X-ray (environmental) properties of the surrounding intracluster medium (ICM).

For the composite case, the relative weights of the two components, as well as their metallicities, are not constrained during the fits. The wavelength range that we consider here (3500 to 4900 \AA{}) contains absorption features, for example the higher-order Balmer lines, that are very sensitive to age but insensitive to metallicity. The region also contains features, for example CN and Fe, that are more sensitive to metallicity and less so to age. The combined measurement of all the spectral features makes accurate age determination possible. The derived stellar population properties are summarised in Table \ref{ages}. 

Our metallicity determinations are not as robustly constrained as we would have liked. The results in Table \ref{ages} imply old stellar populations that are severely metal-poor ([Fe/H] $\sim$ --0.8). This is an artefact of the stellar population models using solar-scaled Padova \citep{Girardi2000} isochrones, which do not predict the right abundances (in particular, the alpha-elements) for very massive old early-type galaxies. In addition, the best metallicity-sensitive features for old elliptical galaxies (certain Fe, Mg absorption lines) fall outside the wavelength range considered here. We do not extrapolate by including corrections from computationally-derived model atmospheres, as this introduces uncertainty. However, the higher-order Balmer lines and other blue absorption features (and hence age determination) are insensitive to metallicity and alpha-elements (see discussion in \citealt{Vazdekis2015}). Thus the poorly-constrained metallicity does not have an influence on the ages or the fractions of young stellar components derived here. A luminosity contribution of a young 100 Myr old stellar component of $\sim$ 50 \% only constitutes $\sim$ 1\% of the mass contribution, so the errors on the derived mass fractions are small.

The four galaxies marked with a $\star$ in Table \ref{ages} all exhibit strong optical emission lines in their spectra in both apertures. 
These also directly correspond to the galaxies where the observed spectra are best represented by a younger ($<$ 2\ Gyr) stellar component superposed on top of an intermediate/old stellar component. Ionised AGN emission, if present, is typically seen only in the inner 5 kpc and should thus be contained within the inner aperture. The presence of emission lines in the outer aperture, as well as the young stellar components, point to star formation as the dominant photo-ionisation mechanism.

\begin{table*}
\begin{tabular}{l c c c r r r c}
\hline
Name & Aperture & Component & Age $\rmn{(Myr)}$ & [Fe/H] dex & Luminosity fraction & Mass fraction & Classification (CCCP)\\
\hline
\multicolumn{8}{c |}{Young stellar components (section 4.1)}\\
\hline
Abell 383$^\star$ & inner & 1 & 100 $\pm$ 50 & 0.20 $\pm$ 0.10 & 12 & $<$ 1& blue\\
& inner & 2 & 4360 $\pm$ 520 & 0.20 $\pm$ 0.05 & 88 & $>$99 &\\
& outer & SSP & 4080 $\pm$ 860 & 0.18 $\pm$ 0.06 & 100 & 100 &\\
Abell 1835$^\star$ & inner & 1 & 200 $\pm$ 100 & 0.20 $\pm$ 0.10 & 71 & 7 & blue\\
& inner & 2 & 15110 $\pm$ 2000 & --1.21 $\pm$ 0.10 & 29 & 93 & \\
&  outer & 1 & 100 $\pm$ 50 & 0.05 $\pm$ 0.35 & 60 & 2 & \\
& outer & 2 & 17000 $\pm$ 1800 & --0.44 $\pm$ 0.07 & 40 & 98 & \\
Abell 2390$^\star$ & inner & 1 & 100 $\pm$ 50 & 0.10 $\pm$ 0.47 & 47& 1 & blue\\
& inner & 2 & 7280 $\pm$ 1220 & 0.20 $\pm$ 0.10 & 53 & 99 &\\
&  outer & 1 & 100 $\pm$ 50 & 0.20 $\pm$ 0.20 & 47 & 1 & \\
& outer & 2 & 5500 $\pm$ 300 & 0.07 $\pm$ 0.10 & 53 & 99& \\
MS 1455+22$^\star$ & inner & 1 & 110 $\pm$ 30 & 0.13 $\pm$ 0.63 & 29 & 1 & blue\\
& inner & 2 & 3440 $\pm$ 680 & 0.16 $\pm$ 0.10 & 71 & 99 & \\
&  outer & 1 & 240 $\pm$ 100 & 0.15 $\pm$ 0.58 & 37 & 3 &\\
&  outer & 2 & 2680 $\pm$ 380 & 0.20 $\pm$ 0.04 & 63 & 97 &\\
\hline
\multicolumn{8}{c |}{Intermediate components (section 4.2)}\\
\hline
Abell 68 & inner & SSP & 4420 $\pm$ 1100 & 0.20 $\pm$ 0.05 & 100 & 100 & blue\\
&  outer & SSP & 4450 $\pm$ 760 & 0.20 $\pm$ 0.10 & 100 & 100 & \\
Abell 209 & inner & SSP & 4430 $\pm$ 1310 & 0.20 $\pm$ 0.05 & 100 & 100 & red\\
& outer  & SSP & 7530 $\pm$ 2440 & 0.19 $\pm$ 0.04 & 100 & 100 & \\
Abell 267 & inner & SSP & 4070 $\pm$ 130 & 0.20 $\pm$ 0.05 & 100 & 100 & red\\
&  outer & SSP & 4010 $\pm$ 200 & 0.20 $\pm$ 0.06 & 100 & 100 & \\
Abell 568 & inner & SSP & 4300 $\pm$ 390 & 0.20 $\pm$ 0.05 & 100 & 100 & red\\
& outer  & SSP & 4160 $\pm$ 900 & 0.15 $\pm$ 0.09 & 100 & 100 &\\
Abell 611 & inner & SSP & 3810 $\pm$ 380 & 0.18 $\pm$ 0.05 & 100 & 100& red\\
&  outer & SSP & 14870 $\pm$ 3600 & --0.43 $\pm$ 0.10 & 100 & 100 &\\
Abell 963 & inner & SSP & 7610 $\pm$ 590 & 0.20 $\pm$ 0.06 & 100 & 100& red\\
& outer  & SSP & 15000 $\pm$ 3250 & --0.23 $\pm$ 0.11 & 100 & 100 &\\
Abell 1763 & inner & SSP & 7410 $\pm$ 460 & 0.20 $\pm$ 0.05 & 100 & 100 & red\\
& outer  & SSP & 7500 $\pm$ 1250 & 0.20 $\pm$ 0.08 & 100 & 100 & \\
Abell 1942 & inner & SSP & 3880 $\pm$ 140 & 0.20 $\pm$ 0.10 & 100 & 100 & red\\
& outer  & SSP & 4200 $\pm$ 1060 & 0.18 $\pm$ 0.07 & 100 & 100 & \\
Abell 2104 & inner & SSP & 5530 $\pm$ 190 & 0.20 $\pm$ 0.04 & 100 & 100 & red\\
& outer & SSP & 4460 $\pm$ 1280 & 0.20 $\pm$ 0.10 & 100 & 100 & \\
Abell 2259 & inner & SSP & 4570 $\pm$ 380 & 0.19 $\pm$ 0.02 & 100 & 100 & red\\
& outer  & SSP & 4510 $\pm$ 710 & 0.20 $\pm$ 0.10 & 100 & 100 & \\
Abell 2261 & inner & SSP & 4410 $\pm$ 560 & 0.20 $\pm$ 0.05 & 100 & 100 & red\\
& outer  & SSP & 4400 $\pm$ 1530 & 0.20 $\pm$ 0.10 & 100 & 100 & \\
Abell 2537 & inner & SSP & 4230 $\pm$ 920 & 0.15 $\pm$ 0.05 & 100 & 100 & red\\
&  outer & SSP & 15720 $\pm$ 5200 & --0.35 $\pm$ 0.10 & 100 & 100 & \\
MS 0440+02 & inner & SSP & 4370 $\pm$ 160 & 0.20 $\pm$ 0.03 & 100 & 100 & red\\
& outer  & SSP & 15400$\pm$ 6280 & --0.88 $\pm$ 0.05 & 100 & 100 & \\
MS 0906+11 & inner & SSP & 4430 $\pm$ 870 & 0.20 $\pm$ 0.07 & 100 & 100 & red\\
& outer  & SSP & 5750 $\pm$ 520 & 0.20 $\pm$ 0.10 & 100 & 100 & \\
\hline
\multicolumn{8}{c |}{Old stellar population (section 4.3)}\\
\hline
Abell 1689 & inner & SSP & 15330 $\pm$ 1410 & --0.28 $\pm$ 0.03 & 100 & 100 & red\\
& outer  & SSP & 14250 $\pm$ 2310 & --0.34 $\pm$ 0.07 & 100 & 100 &\\
\hline
\end{tabular}
\caption{Derived luminosity-weighted, stellar population properties for the BCGs. The young (1) and old components (2) are indicated for the four BCGs for which composite populations were needed to fit the observed spectra. The same four BCGs, marked with $\star$, exhibit emission lines in their optical spectra. The last column indicates the colour of the core as measured by \citet{Bildfell2008}. The error arrays, described in Section 2, are used to derive the errors on the stellar population properties, as described in Section 3.}
\label{ages}
\end{table*}

\begin{figure}
   \centering
   \includegraphics[scale=0.32, trim=3mm 12mm 3mm 3mm, clip]{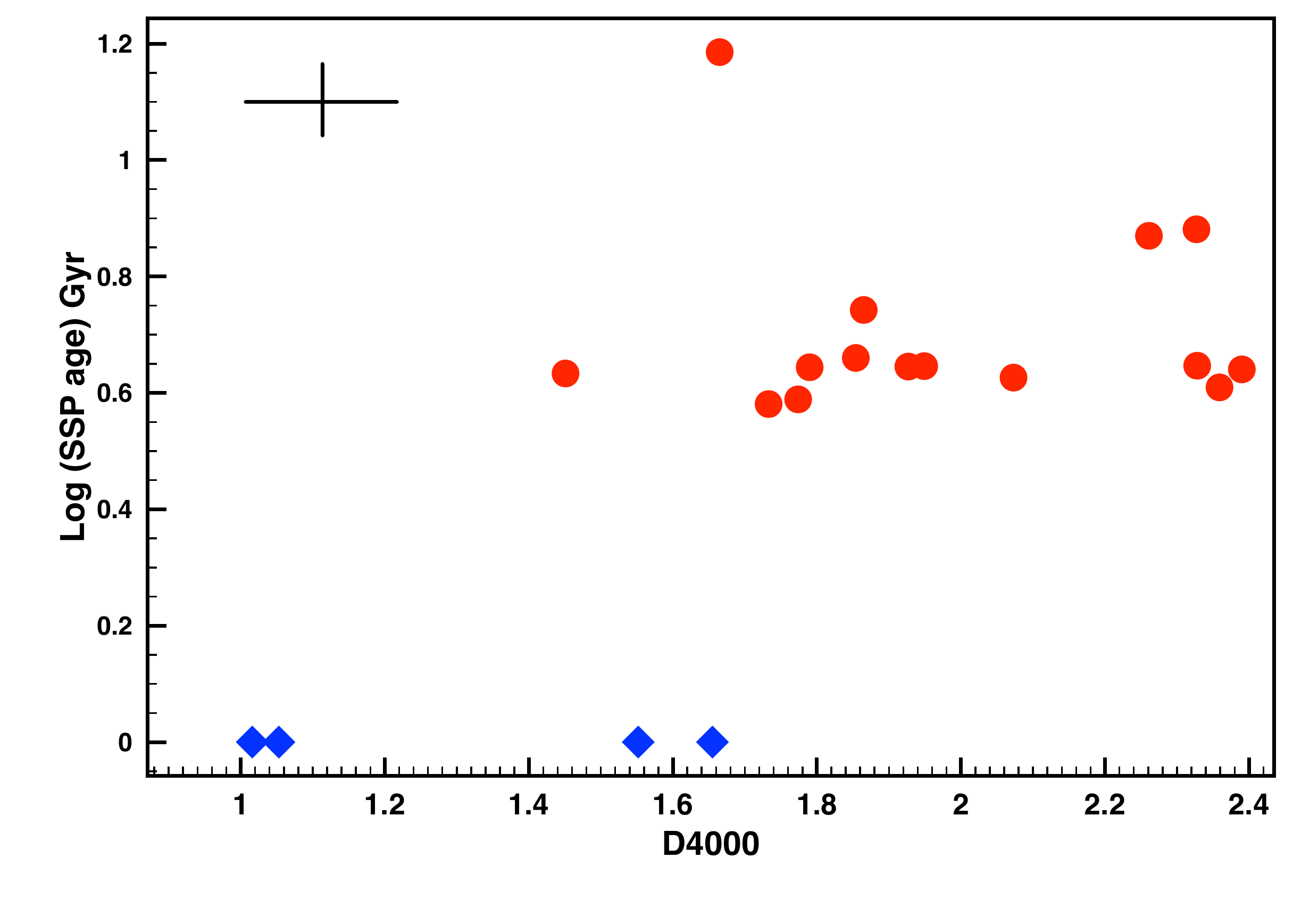}
   \caption[]{D4000 index measurements compared to the age measurements with full-spectrum fitting. Typical error bars are indicated in the upper left corner. It shows the bimodality ($\sim 1.6$) between old and young galaxies.} 
   \label{d4000}
\end{figure}

The strength of the D4000 index also offers a useful diagnostic separating old, red galaxies and star forming galaxies that can be complimentary to the full-spectrum fitting. The D4000 index increases steadily with age, and can be used to identify galaxies with star formation in the last Gyr \citep{Kauffmann2003, Brinchmann2004}. We use the narrow-band D4000 definition of \citet{Balogh1999} as it is less prone to be influenced by reddening. Dust along the line of sight will attenuate the blue light, and therefore change the shape of the observed spectrum. A D4000 index measurement of less than 1.6 is a good indication of a galaxy with a recent star formation episode, and larger than 1.6 is consistent with old galaxies. Even though the errors on the individual index measurements are considerably higher than for full-spectrum fitting, we find consistent results in that the four galaxies with young components also have some of the lowest D4000 measurements (Figure \ref{d4000})\footnote{We emphasise that the four galaxies with prominent young components were not accurately fitted with SSPs, but with a composite population. They are therefore plotted at Log (SSP age) = 0 to facilitate comparisons with the X-axis quantities. These \textit{ad hoc} SSP ages should not be directly compared to the SSP ages plotted on the Y-axis.}.

\section{Discussion}

Even though the BCGs in the sample were selected to be close to the peak of X-ray luminous clusters, they show a large variety in the derived stellar populations, and most probable SFHs. We broadly separate and discuss the different stellar populations below.

\begin{figure*}
   \centering
   \mbox{\subfigure{\includegraphics[scale=0.52, trim=20mm 140mm 20mm 30mm, clip]{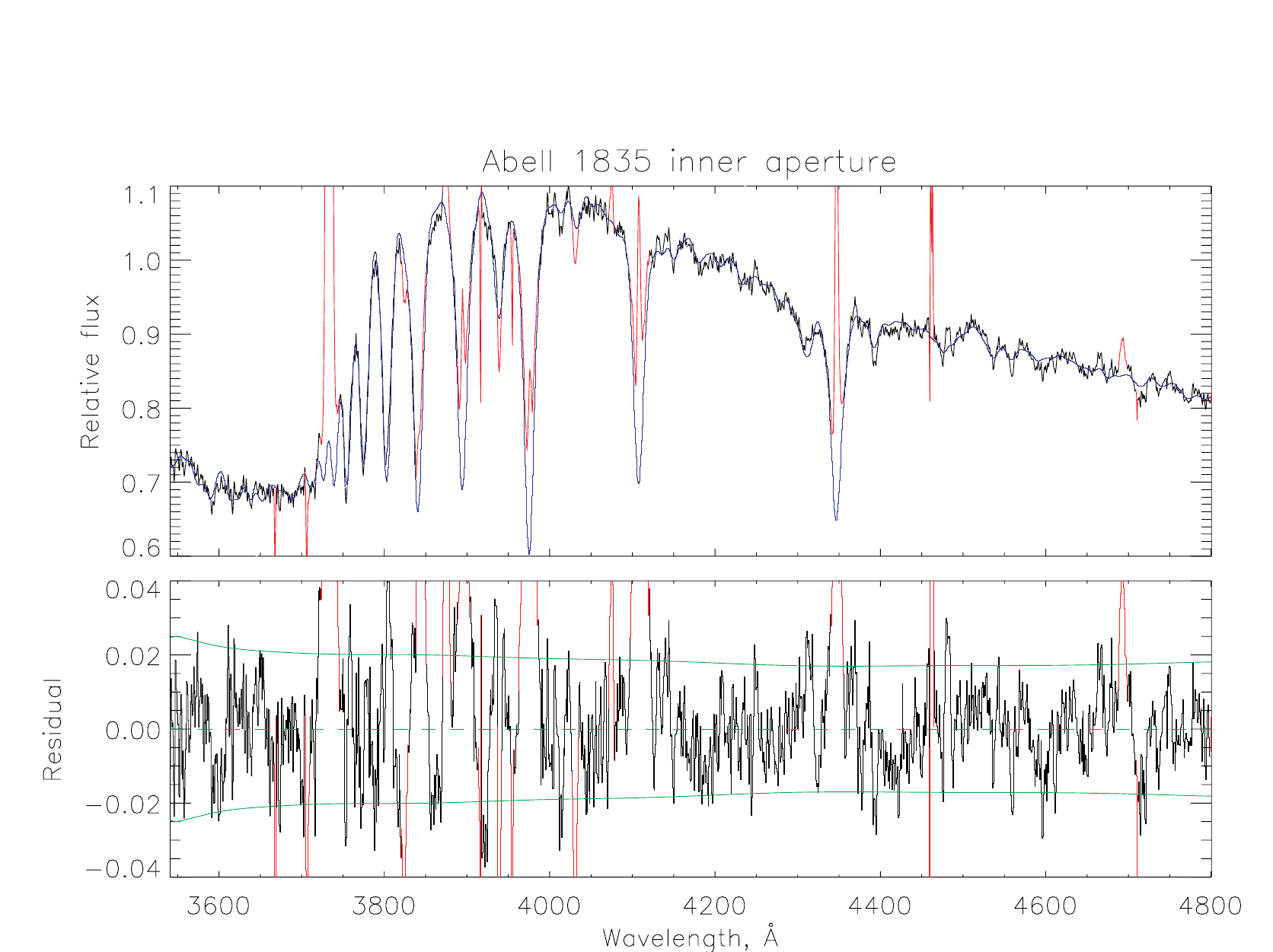}}\quad
  \subfigure{\includegraphics[scale=0.52, trim=20mm 140mm 20mm 30mm, clip]{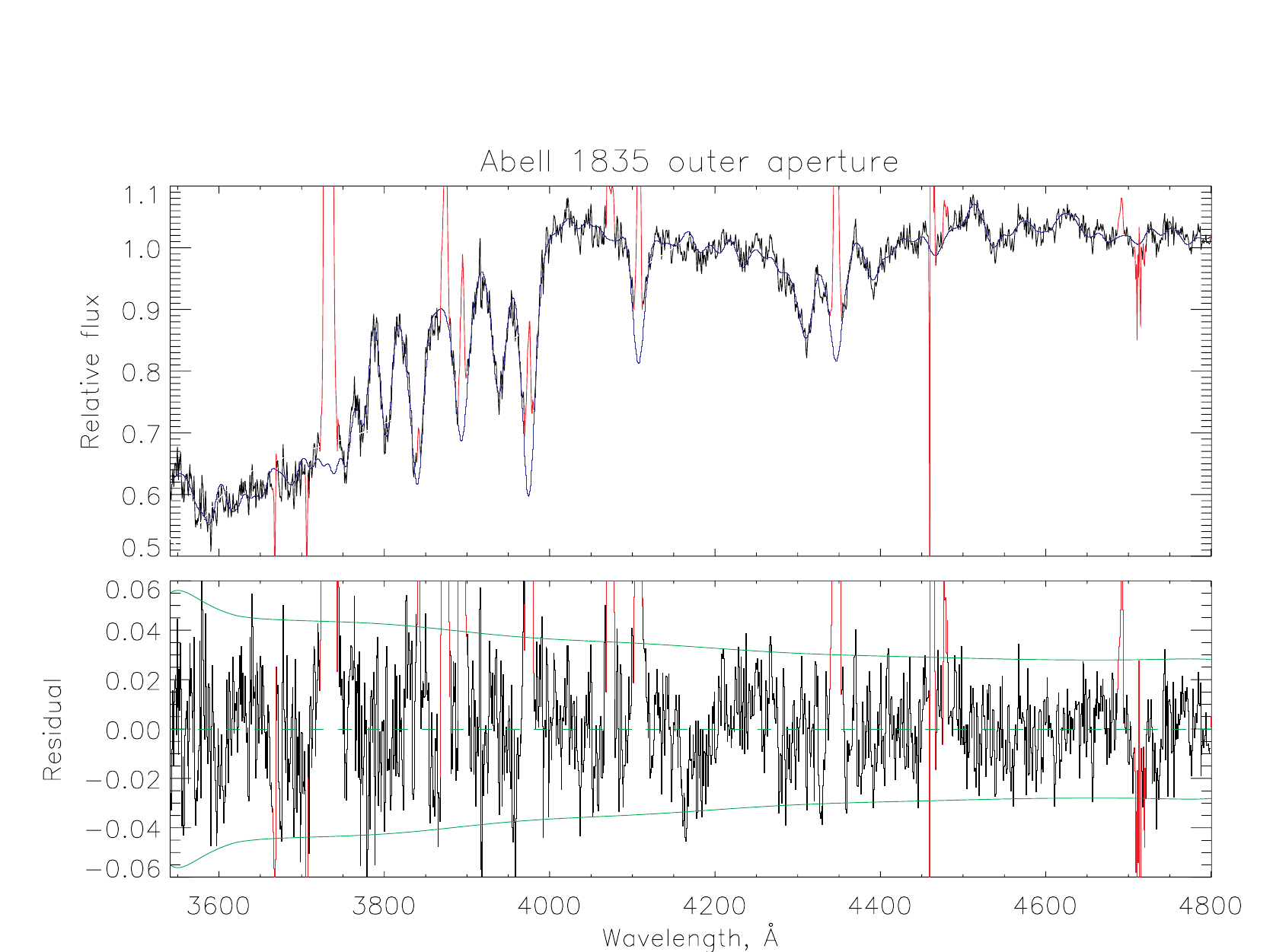}} }
   \mbox{\subfigure{\includegraphics[scale=0.52, trim=20mm 140mm 20mm 30mm, clip]{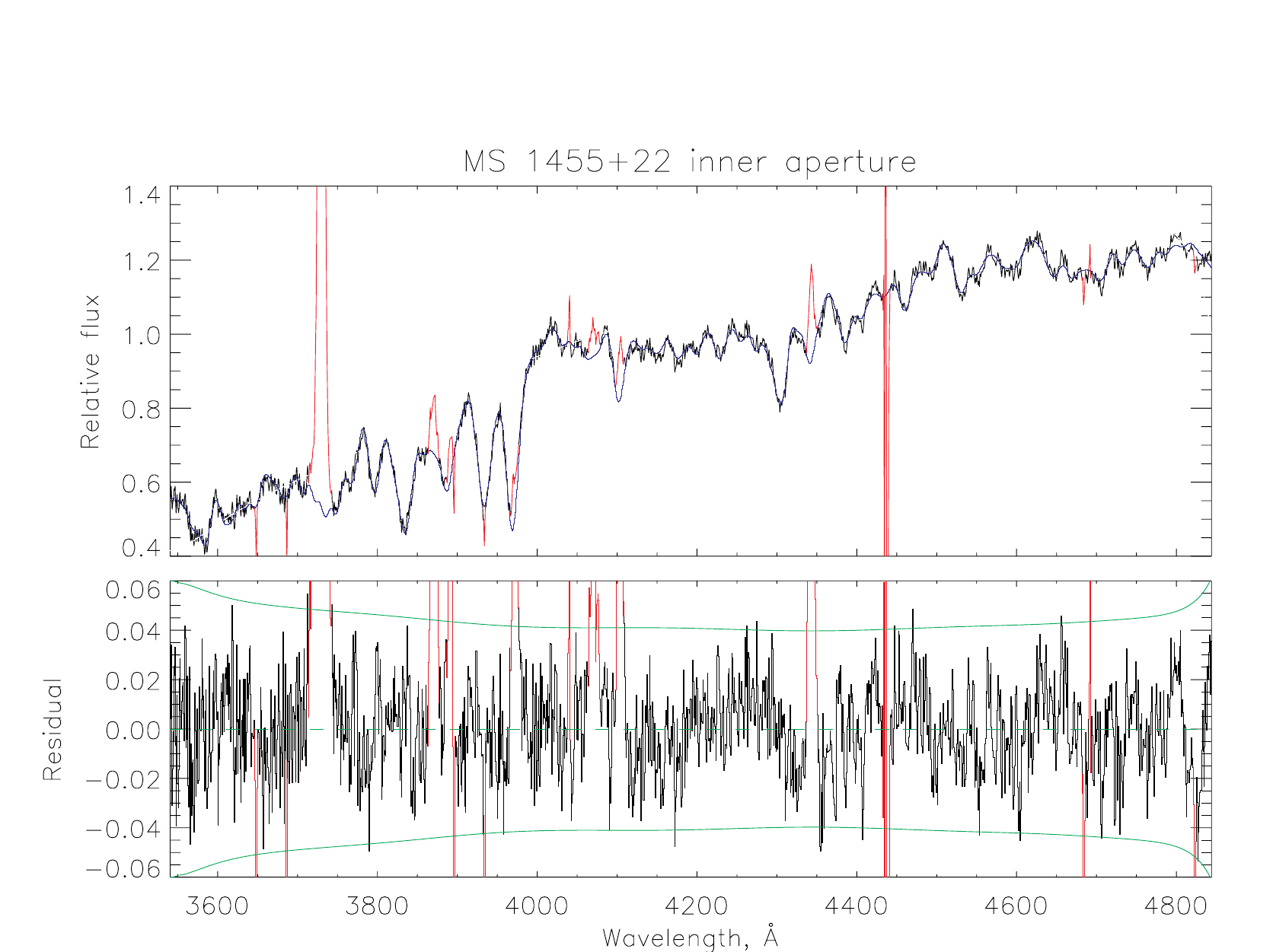}}\quad
  \subfigure{\includegraphics[scale=0.52, trim=20mm 140mm 20mm 30mm, clip]{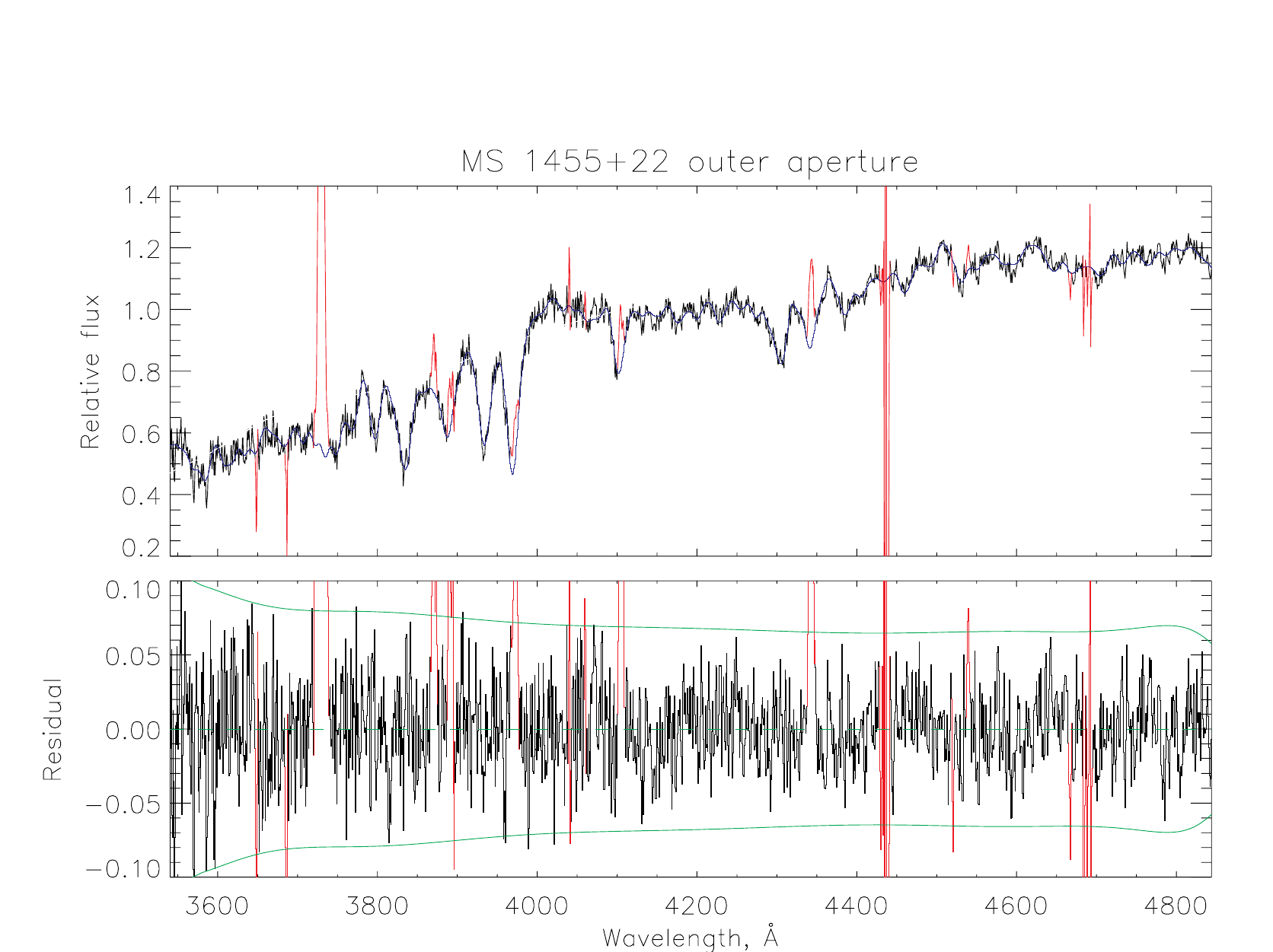}} }
   \mbox{\subfigure{\includegraphics[scale=0.52, trim=20mm 140mm 20mm 30mm, clip]{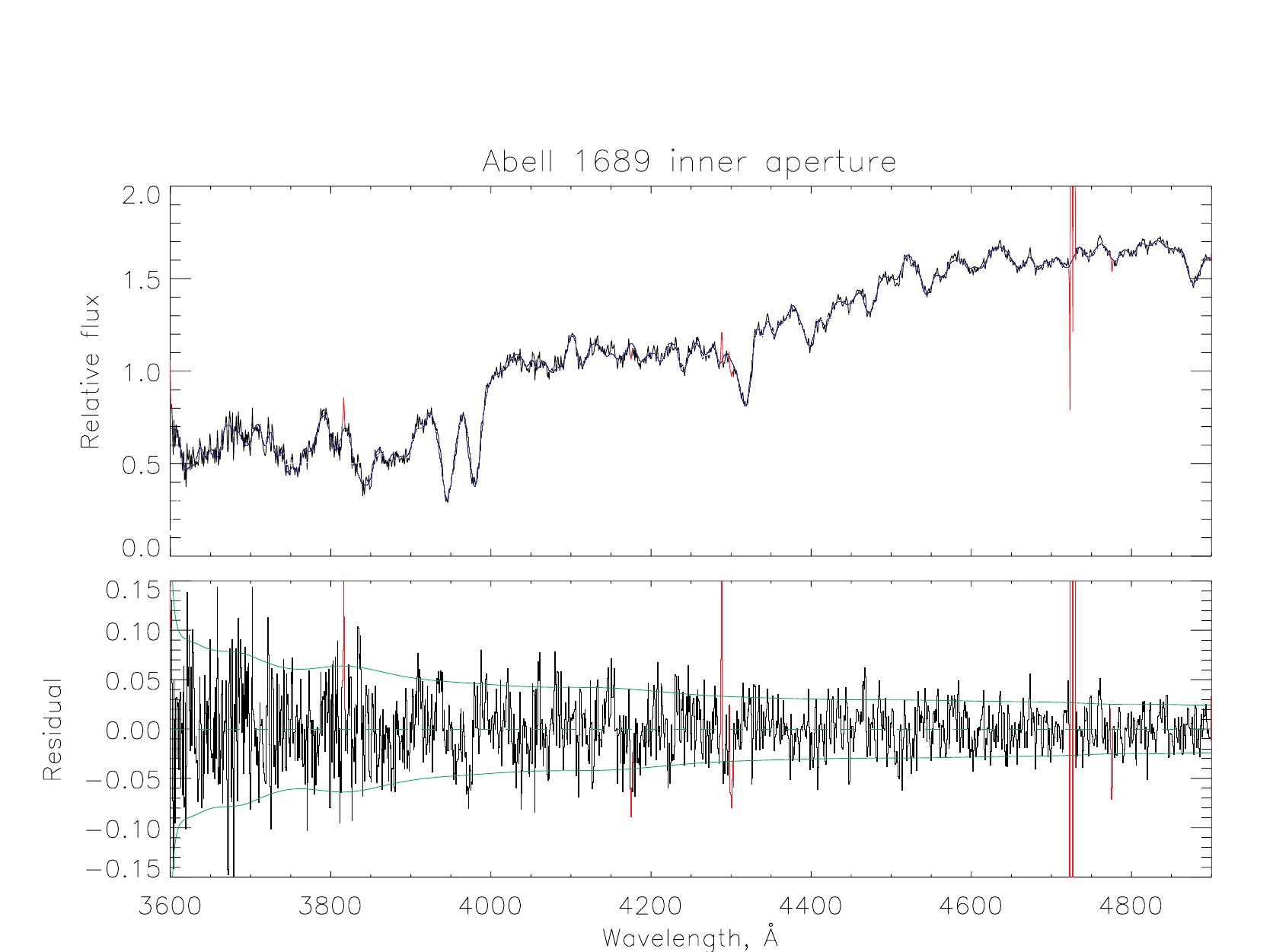}}\quad
  \subfigure{\includegraphics[scale=0.52, trim=20mm 140mm 20mm 30mm, clip]{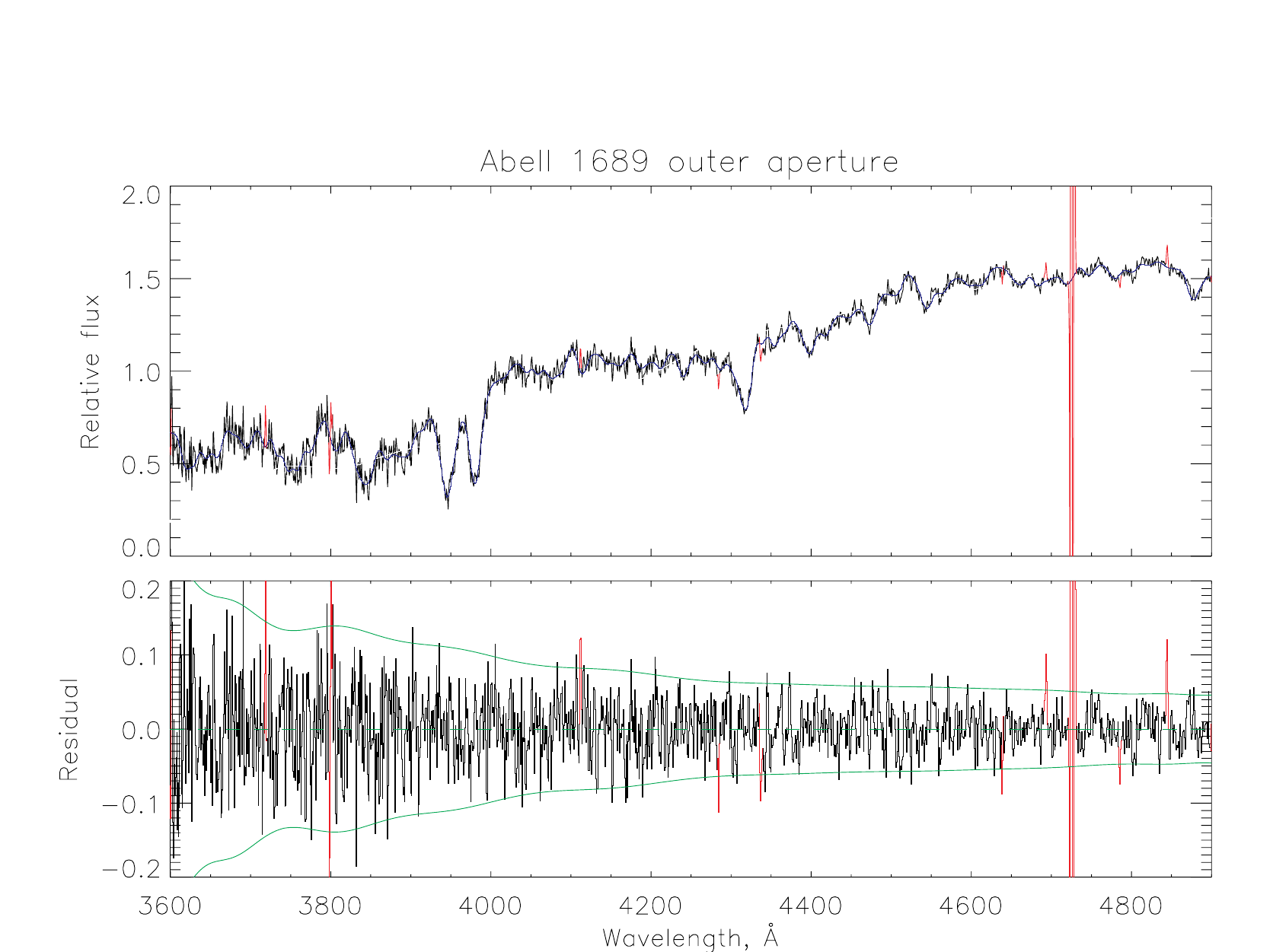}} }
   \caption[]{Upper: the fit to the inner (left) and outer (right) aperture of Abell 1835, an example of vigorous recent star formation. Middle: the fit to the inner (left) and outer (right) aperture of MS 1455+22, an example of moderate recent star formation. Lower: the fit to the inner (left) and outer (right) aperture of Abell 1689, an example of an old, passively evolving BCG. The residuals are shown in the bottom panel, along with the 1$\sigma$-error (calculated from the error spectra which include Poisson as well as systematic noise) in green. The emission lines (red) were masked to fit the underlying absorption features.}
   \label{fit}
\end{figure*}

\subsection{Prominent young stellar components}
\label{youngcomponents}

The presence of a younger component in the stellar populations is easily identified as it exhibits prominent spectral features and represents a large fraction of the luminosity. The BCGs that show these young populations (in Abell 383, Abell 1835, Abell 2390 and MS 1455+22) are in qualitative agreement with the detection of optical photometric ``blue" cores by \citet{Bildfell2008} in some of the BCGs in the CCCP sample (see Table \ref{ages}), with the exception of the BCG in Abell 68, which is indicated as having a blue core in \citet{Bildfell2008} but has no optical emission lines in the spectra, and is very well approximated with an SSP. This galaxy was noted to be an exception to the general trend by \citet{Bildfell2008} because it is located in a cluster with a central entropy, $K_{0}$, well above the 30 keV cm$^{2}$ threshold (BCGs in clusters with high central entropies generally do not show evidence of recent star formation episodes, see section \ref{clusters}). The sizes of the blue cores (typically 20 to 30 kpc), as measured by \citet{Bildfell2008}, in the four galaxies with detectable young stellar components also qualitatively agree with our stellar population fits in the different apertures.

The young component of the BCG in Abell 1835 is particularly dominant in the luminosity fraction, to such an extent that the fit to the older, less luminous, component is uncertain. The spectrum shows clear features of A-type stars (Figure \ref{fit}), which is very unusual for a massive central cluster elliptical galaxy. The BCG in Abell 383 is also unusual in that it shows optical emission lines in the inner and outer aperture, even though the outer aperture is best approximated by an SSP.

Our results constraining the mass fraction of the young component agrees with, and are complimentary to, the SFRs derived in the UV (see the detailed comparison to UV observations of \citet{Pipino2009} and \citet{Donahue2015} in section \ref{multiwavelength}). 

All four young stellar components that we detect are $\sim$ 100 -- 200 Myr old. This can signify a single burst of star formation less than 200 Myr ago, or repeated bursts of which the most recent was in the last 200 Myr. In case of the latter, the spectral features and luminosity contribution of the most recent burst are much more prominent than those of preceding episodes, and the age of the young component will converge on 100 -- 200 Myr during the stellar population analysis.

If there is no very young ($\sim 200$ Myr) component, but there is a component with age still below 1 Gyr, then we would have been able to detect it. As a demonstration, we construct a set of mock spectra consisting of a 1 to 5\% mass fraction contribution from either a 200, 400, 800, 1200, 1600, or a 2000 Myr stellar population (thus building a grid of 30 spectra) superposed on a 15 Gyr population. We use the same stellar population models and IMF as in our analysis of the observations. We use Monte-Carlo simulations (where n = 200) to fully incorporate the observational errors in the spectra, which has a non negligible influence on the ability to detect small fractions of young populations. We take the signal in each spectral bin and scatter it up or down within observational errors (S/N = 40) and take each realization as a possible spectrum for analysis, and convolve the mock spectra with a Gaussian of width 250 km s$^{-1}$ to account for the velocity dispersion of the BCGs.ÊWe keep the metallicity of the stellar components fixed at typical metallicities for components of that age (e.g. [Fe/H] = --0.5 for a 15 Gyr population; [Fe/H] = 0.2 for a 200 Myr population), and we analyse all the mock spectra in the exact manner in which we analyse the observational data. We use the SSP as well as composite model fits for each of the 200 realizations of the 30 spectra to asses the ability to recover the small fraction of the young stellar populations. We do so by measuring the deviation from the mean SSP age, and the deviation from the expected CSP young age.

The resulting grid (Table \ref{Grid}) contains the recovered ages as well as the errors on the ages of the SSP and composite model fits, and is an accurate representation of our ability to detect the young populations within the observational errors. It illustrates that: 1\% to 5\% of the 200, 400, 800 Myr populations can be reliably detected using a composite model fit. Only the larger of the 1200 Myr (4\% and 5\%) and 1600 Myr (5\%) bursts can be reliably detected using a composite model. None of the five 2000 Myr old populations (as well as the smaller mass fractions of 1200 and 1600 Myr populations) can be reliably detected with a composite model, and an SSP fit is the preferred fit as dictated by our analysis method. Thus, even bursts as small as 1\% in mass as far back as 800 Myrs ought to have been detectable. Only bursts that contribute $>$ 4\% in mass ought to have been detectable if they occurred 1 -- 2 Gyrs ago.

\begin{table*}
\begin{tabular}{l c c c c c}
\hline
\multicolumn{6}{c}{Age and mass fraction contribution of young stellar component}\\
 & 1\% & 2\% & 3\% & 4\% & 5\% \\
\hline
\multicolumn{6}{c}{200 Myr}\\
SSP	& 4550 $\pm$ 200 & 3066 $\pm$ 93 & 1602 $\pm$ 47 & 1324 $\pm$ 25 & 1160 $\pm$ 19\\
Composite & \textbf{208 $\pm$ 35} & \textbf{205 $\pm$ 19} & \textbf{208 $\pm$ 16} & \textbf{205 $\pm$ 13} & \textbf{206 $\pm$12}\\
\multicolumn{6}{c}{400 Myr}\\
SSP	& 6699 $\pm$ 402 & 3994 $\pm$ 300 & 3078 $\pm$ 114 & 2770 $\pm$ 157 & 1675 $\pm$ 90\\
Composite & \textbf{394 $\pm$ 154} & \textbf{369 $\pm$ 54} & \textbf{368 $\pm$ 40} & \textbf{368 $\pm$ 32} & \textbf{377 $\pm$ 30}\\
\multicolumn{6}{c}{800 Myr}\\
SSP & 12085 $\pm$ 937 & 9618 $\pm$ 854 &	8659 $\pm$ 753 & 7778 $\pm$ 534 & 6273 $\pm$ 358\\
Composite & \textbf{1396 $\pm$ 760} & \textbf{1138 $\pm$ 315} &	\textbf{1005 $\pm$ 217} & \textbf{919 $\pm$ 180} & \textbf{819 $\pm$ 142}\\
\multicolumn{6}{c}{1200 Myr}\\
SSP & \textbf{13444 $\pm$ 831} & \textbf{11863 $\pm$ 	911} & \textbf{10619 $\pm$ 918} & 9885 $\pm$ 893 & 9383 $\pm$ 837\\
Composite & 3094 $\pm$ 9893 & 2495 $\pm$ 4840 & 2378 $\pm$ 5336 & \textbf{1698 $\pm$ 574} & \textbf{1591 $\pm$ 339}\\
\multicolumn{6}{c}{1600 Myr}\\
SSP	& \textbf{13570 $\pm$ 851} & \textbf{12608 $\pm$ 861} & \textbf{12321 $\pm$ 869} & \textbf{11482 $\pm$ 882} & 10944 $\pm$ 905\\
Composite & 3416 $\pm$ 3226 & 3428 $\pm$ 2854 & 2403 $\pm$ 1340 & 2839 $\pm$ 1112 & \textbf{1612 $\pm$ 759}\\
\multicolumn{6}{c}{2000 Myr}\\
SSP & \textbf{14187 $\pm$ 874} & \textbf{13627 $\pm$ 859} & \textbf{12924 $\pm$ 901} & \textbf{12365 $\pm$ 861} &	 \textbf{11786 $\pm$ 874}\\
Composite & 2409 $\pm$ 5894 & 3520 $\pm$ 5312 & 3463 $\pm$ 3123	& 3029 $\pm$ 2068 & 2606 $\pm$ 1844\\								
\hline
\end{tabular}
\caption{Grid illustrating the ages and errors of the best fits (SSP vs. composite models) for mock spectra of different mass fraction contributions of a young stellar population superposed on a 15 Gyr population. The best fits (lowest $\chi^{2}$ and uncertainty) are indicated in bold.}
\label{Grid}
\end{table*}

Looking at the mass accretion histories suggested by the recent hydrodynamical simulations of the ``cold mode accretion model'' \citep{Li2015, Prasad2015}, we find that typically the deposition of cold gas onto the central BCGs occurs via groups of episodic events occurring in quick succession, with each individual event expected to give rise to a starburst.Ê This, in turn, led us to investigate whether it is possible to detect differences in the composite spectra if the young population was formed in a recent single or a sequence of multiple bursts (spread out over the last Gyr).Ê After all, stellar population analysis assigns a most probable, but very rarely a unique, SFH to a galaxy.

\begin{figure*}
   \centering
   \mbox{\subfigure{\includegraphics[width=19cm, height=10.5cm, trim=20mm 140mm 10mm 40mm, clip]{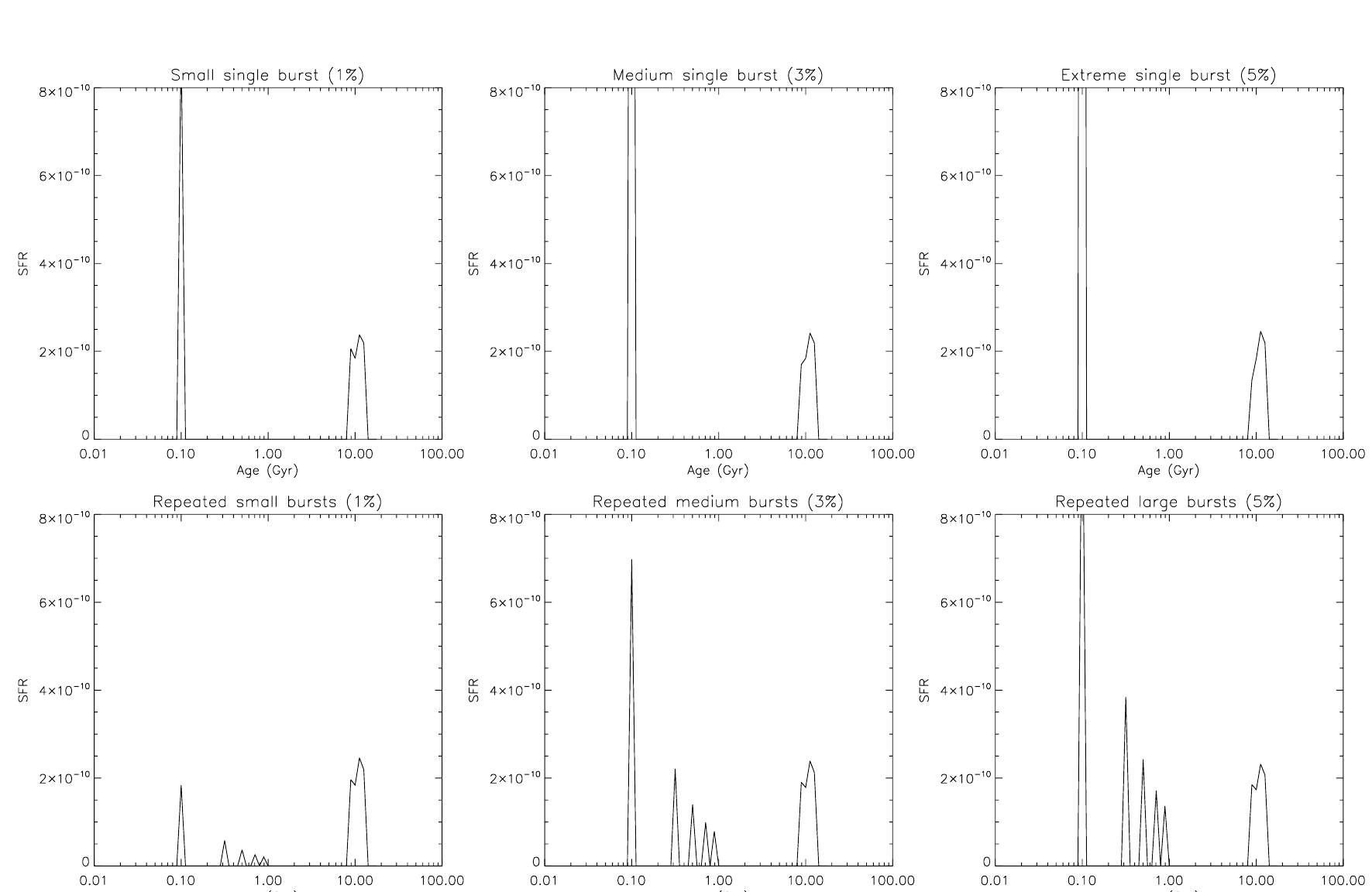}}}
  \mbox{\subfigure{\includegraphics[width=19cm, height=10.5cm, trim=20mm 140mm 10mm 40mm, clip]{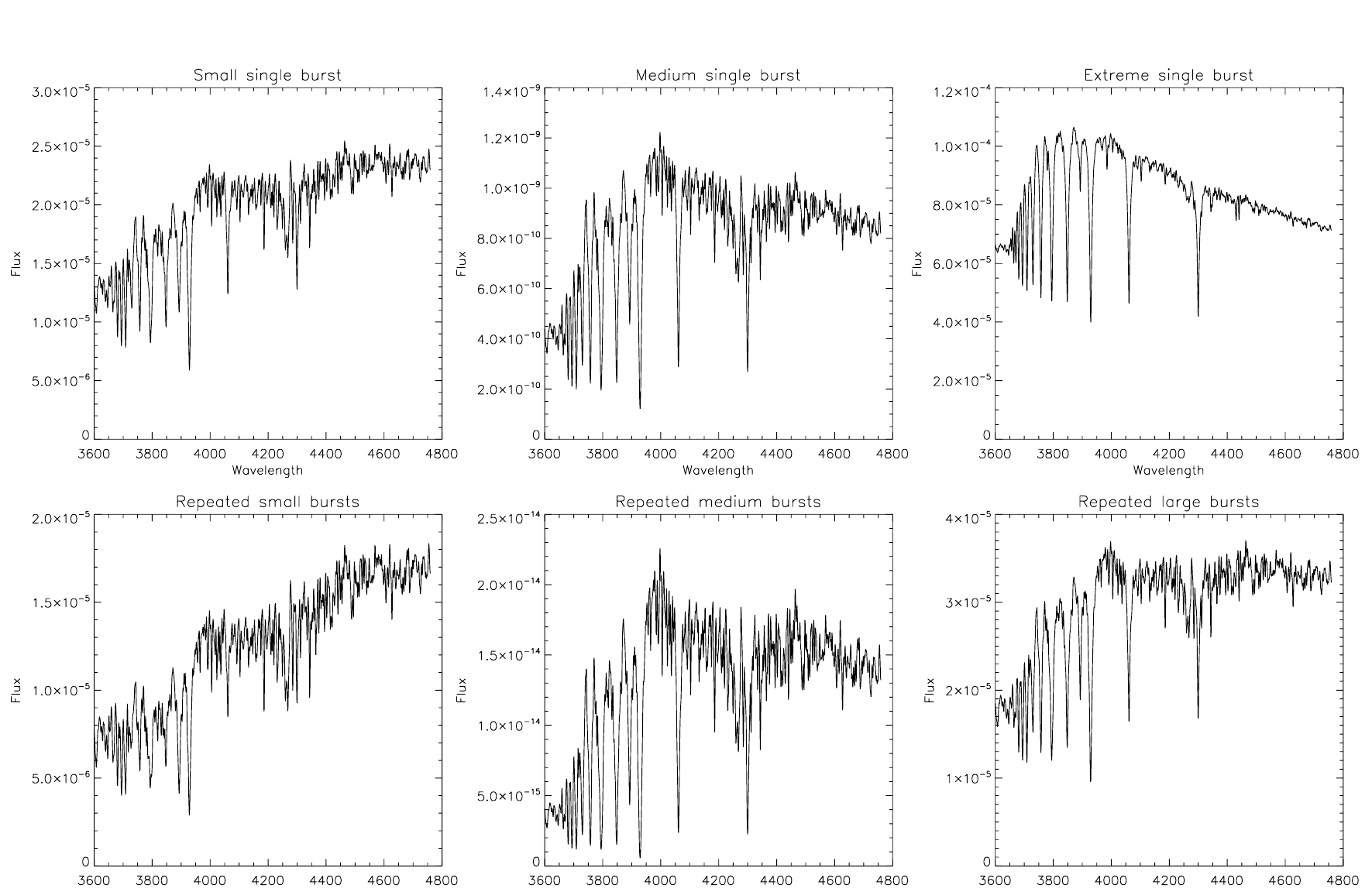}}}
   \caption[]{Composite spectra expected for broadly different star formation history scenarios with at least one episode of star formation during the last Gyr. The different scenarios are a single recent burst (forming 1 per cent of the total stellar mass), small repeated bursts (forming 1 per cent of the mass), a larger single recent burst (forming 5 per cent of the mass), and repeated bursts (forming 5 per cent of the mass). The SFR is given as M$_{\odot}$/yr normalised to 1 M$_{\odot}$.}
   \label{SFHs}
\end{figure*}

Using the same stellar population models used in the preceding analyses, we construct composite spectra corresponding to the case where the young population is formed via a single recent burst, or via a sequence of bursts (every 200 Myrs over the past Gyr).ÊÊ We prepare these spectra in the same manner as described above and again, analyze these in the exactly as we analyse the observational data.ÊÊÊ In Figure \ref{SFHs}, we show an example where the young component comprises 1, 3 and 5\% of the total stellar mass.Ê We find that if the star formation episode(s) are small (cumulatively contributing less than $\sim$ 3\% of the stellar mass in the last Gyr), then there are no discernible features that will identify previous (repeated) bursts beyond the most recent (and the luminous) star formation episode.Ê Only if the episode(s) are large (cumulatively contributing more than $\sim$ 5\% of the total stellar mass in the last Gyr), do the predicted spectra corresponding to the single and repeated bursts show prominent detectable differences.ÊÊÊ We note, however, that the above conclusion is specific to the Salpeter IMF.ÊÊ In Section 4.4, we examine the implications of an IMF that is more bottom-heavy than Salpeter.

\subsection{Intermediate components}

Galaxies where the observed spectra are best represented by an SSP model of an intermediate age are the most common in this sample (Table \ref{ages}), but also the most difficult to assign an unique star formation history to.

In general, the SSP ages determined for the inner and outer apertures agree within the error bars; however, there are four BCGs (Abell 611, 963, 2537, MS 0440+02) that shows significantly older luminosity-weighted SSP ages in their outer bins compared to the inner bins (all labelled red cores in \citealt{Bildfell2008}). Age gradients for local BCGs are shallow on average \citep{Loubser2012}, but can show remarkable scatter that are reflective of the BCG accretion history \citep{Oliva2015}. It is likely that the intermediate aged galaxies, and in particular the four BCGs with large age gradients, have had active evolutionary paths and possibly bursts of star formation over cosmic time, generating one or more younger (but older than 2 Gyr) stellar population components superposed on an old population. This is especially likely for the four BCGs (Abell 611, 963, 2537, MS 0440+02) where the outer apertures indicate that the majority of the stars are $\sim 14$ Gyr old, and where all four host clusters have cooling times below $\sim 2$ Gyr (see Figure \ref{Gradtc}). 

We have conducted different tests on the stellar population fits to the above-mentioned galaxies. We found that even though the galaxy might have experienced a complex evolutionary path that differs from passive evolution, an SSP is still the preferred fit as the extraction of different age components that are closer to each other in age (and older than 2 Gyr) from the integrated spectra is often not reliable because of the similarity in the integrated stellar spectra of older populations.

\begin{figure}
   \centering
   \includegraphics[scale=0.47, trim=25mm 130mm 20mm 38mm, clip]{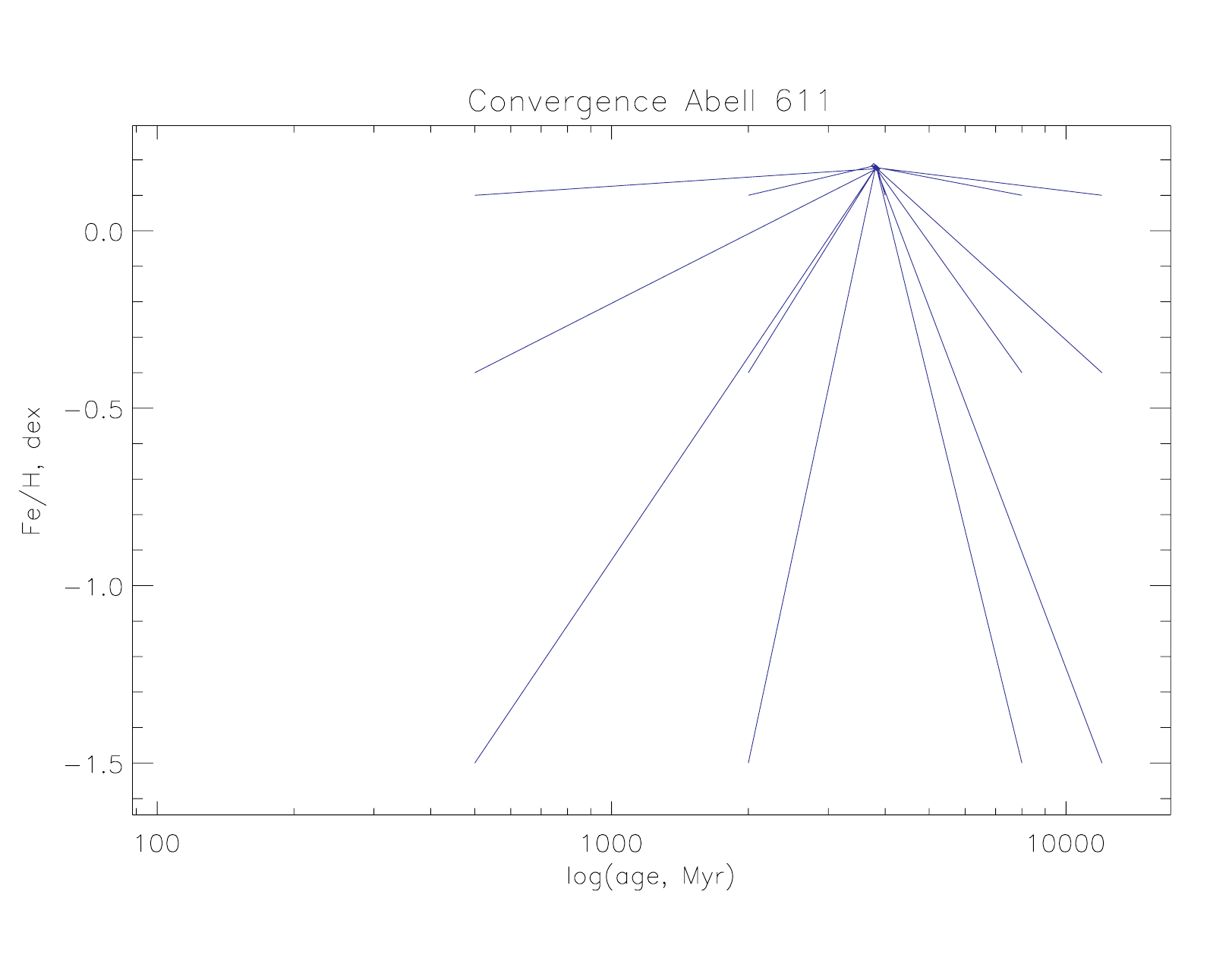}
   \caption[]{An example of how the SSP fit converged for the inner component of Abell 611. Different starting estimates were given for the age and metallicity and the fit converged on the solution presented in Table \ref{ages}.}
   \label{SSPconvergence}
\end{figure}

We show an example of how the SSP fit converges, for different initial estimates for the age and metallicity, for the inner component of Abell 611 in Figure \ref{SSPconvergence}. The SSP fits are a reliable luminosity-weighted estimate of the stellar population age, and will be heavily biased towards an intermediate age if even a small younger/intermediate population is present.

Focusing on the spread in the SSP ages for the inner apertures of the intermediate component galaxies (from Table \ref{ages}), we find that the intrinsic scatter in the ages dominates the statistical errors on the ages. The observed root mean square (rms) of log(age) is found to be 0.094, whereas we estimate a dispersion of 0.051 from the statistical uncertainties in the log of the SSP ages. Under the assumption that these follow a normal distribution in log(age), we obtain an intrinsic rms of 0.079 in log(age). This intrinsic scatter suggests a variety in the evolutionary histories of the intermediate galaxies.


\bigskip

\subsection{Old, ``red $\&$ dead" BCG}

Only one BCG (in Abell 1689) best resembles an old, entirely passively evolving population (Figure \ref{fit}). This galaxy is also recorded as a ``red" core in \citet{Bildfell2008}, and is the only BCG in this X-ray luminous cluster sample that is unlikely to have had star formation episodes during its evolution. 

\subsection{Effect of uncertainty in the IMF and metal abundances}

\begin{figure}
   \centering
   \includegraphics[width=7cm, height=12cm, trim=20mm 130mm 101mm 30mm, clip]{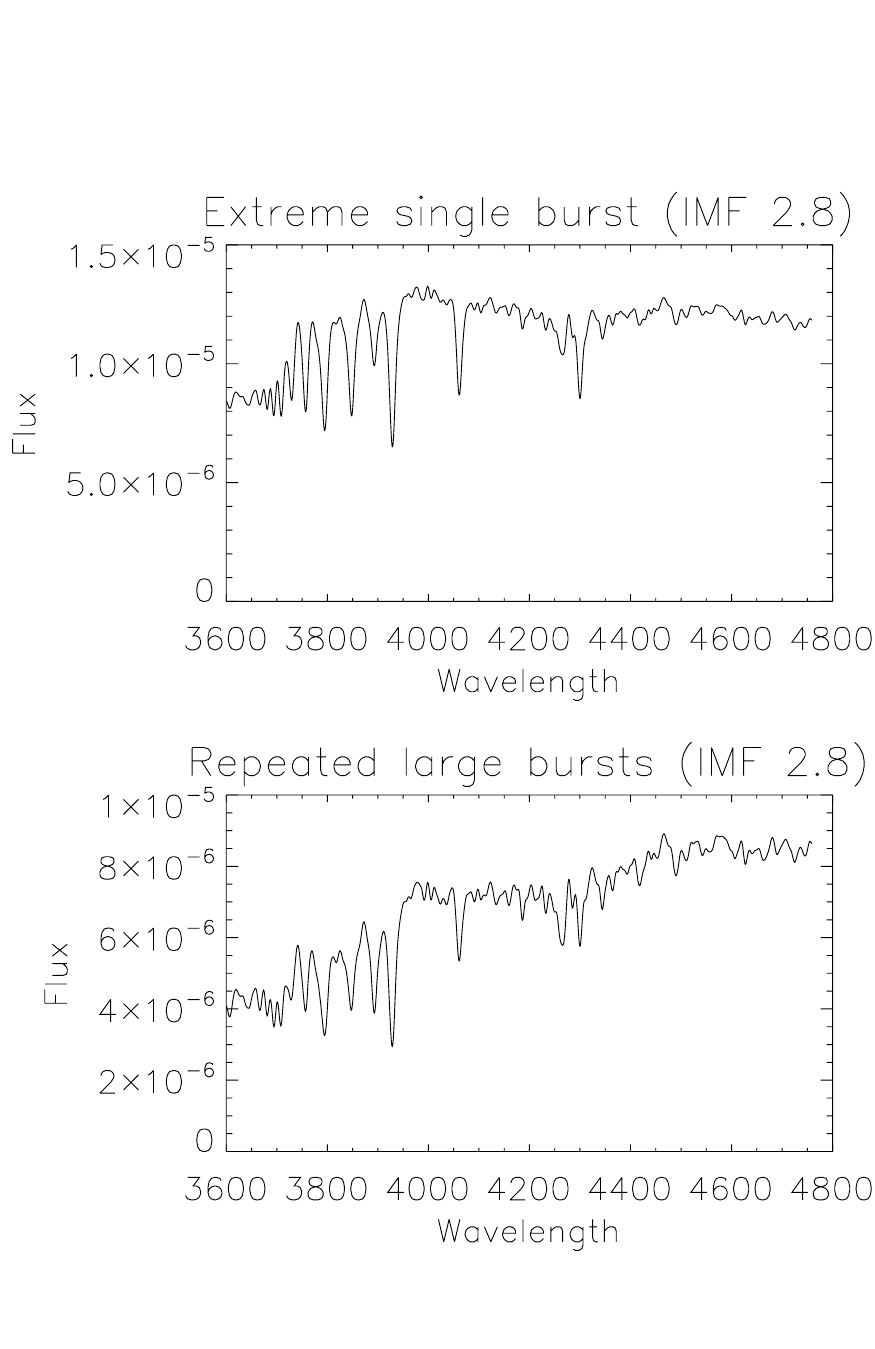}
   \caption[]{Composite spectra expected for different SFH scenarios where there is at least one episode of star formation during the last Gyr, assuming a very bottom-heavy IMF. The different scenarios are a larger single recent burst (forming 5 per cent of the mass), and repeated bursts (forming 5 per cent of the mass). The SFR is given as M$_{\odot}$/yr normalised to 1 M$_{\odot}$. If the IMF in star forming BCGs is more bottom-heavy than a Salpeter IMF, then it is much more difficult to distinguish multiple star formation episodes in the last Gyr from one recent, larger star formation episode (compared to a Salpeter IMF in the bottom-right panels in Figure \ref{SFHs}).}
   \label{IMF}
\end{figure}

Recent observational evidence suggest a variable IMF that becomes progressively more bottom-heavy with mass in early-type galaxies \citep{Conroy2012}. We have repeated the tests described in section 4.1, where we built composite spectra given a particular star formation history, but using a very bottom-heavy (slope 2.8) IMF instead of a Salpeter IMF (slope 2.3). An example, where 5 per cent of the total stellar mass is formed in the last Gyr, is shown in Figure \ref{IMF}. The differences between a single and multiple burst scenario are now less pronounced. Now in order to recover a spectrum resembling that of the inner aperture of Abell 1835 (Figure \ref{fit}), a very recent burst with a mass contribution of more than 10 per cent is required. This is much larger than star formation rates observed in the UV (section \ref{multiwavelength}, \citealt{Pipino2009, Donahue2015}), or the bursts predicted by the hydrodynamical simulations \citep{Li2015, Prasad2015}. Our main conclusion from these tests is therefore: if the IMF in star forming BCGs is more bottom-heavy than a Salpeter IMF, then it is much more difficult to distinguish multiple star formation episodes in the last Gyr from one recent, larger star formation episode.

The large central galaxies in rich clusters are very metal-rich in general (see \citealt{Loubser2009} for stellar populations of nearby BCGs), and the measured metallicity is frequently limited by the upper limit of where the empirical stellar population models ([Fe/H] = 0.2 dex in Vazdekis/MILES models using solar-scaled Padova \citep{Girardi2000} isochrones) are considered reliable \citep{Vazdekis2010}. As mentioned in section 3, we do not extrapolate by including corrections from computationally-derived model atmospheres, as this introduces uncertainty and will not influence the detection of a younger stellar component. Our stellar population results, measured from the blue part of the spectrum and the higher-order Balmer lines, are insensitive to variations in super-solar abundances, like [$\alpha$/Fe] \citep{Vazdekis2015}.\ \citet{Kauffmann2003} (their Figure 3) show that in galaxies with young stellar populations (\ie galaxies with D4000 $<$ 1.6), the difference between the measurements of blue features, like H$\delta$ and D4000 measurements on an index-index plot is significantly larger between models of recent starbursts and models of continuous star formation, than between the same models at different metallicities ($Z_{\odot}$ and 2.5$Z_{\odot}$).

\section{Analysis}
\label{clusters}

\subsection{Comparison with previous results from UV data}
\label{multiwavelength}

Our results for the subsample (Abell 1835, Abell 1942 and Abell 2390) that overlaps with the sample that \citet{Pipino2009} examined via a joint analysis of near-ultraviolet (NUV, from \textit{GALEX}) and optical photometric data agree very well with their findings, including their characterisation of ``blue cores", which they detect in Abell 1835 and Abell 2390 in which we also find young components, but not in Abell 1942 where we do not find a young component. Similarly, our findings for the subsample of BCGs (Abell 383, Abell 209, Abell 2261, Abell 611) also agree with the results of the UV analysis by \citet{Donahue2015}. They find a UV excess possibly associated with star formation in Abell 383, where we also find a young component, but not in the three other BCGs, where we too do not find young components. 

As mentioned in section \ref{youngcomponents}, all the young stellar components that we detect in the BCGs contain stars that are no more than ~200 Myr old, and we ought to have been able to detect stellar populations with a minimum age between 200 Myr and 1 Gyr in the other systems, if present. \citet{Pipino2009} also found that the recent star formation in blue core BCGs typically had an age less than 200 Myr. They suggest that these systems have been repeatedly forming stars for most of their lifetime, which is more plausible than catching these systems all within 200 Myr of a single burst of star formation. In addition, if the latest generation of stars were $\sim 1$ Gyr, then we would expect to see some BCGs in cool-core systems showing `E+A' -type characteristics in their optical spectra.

\citet{Pipino2009} estimate the star formation history of each galaxy in their sample by comparing the NUV and optical photometry to a library of synthetic photometry, generated using a large collection of model histories. They have found Abell 1835 to be star forming at 120 M$_{\odot}$\ yr$^{-1}$, and Abell 2390 at a much lower 5 M$_{\odot}$\ yr$^{-1}$. These estimates are qualitatively in excellent agreement with our findings of an unusually high mass fraction of young stars in Abell 1835 (Figure \ref{fit}), and a much lower mass fraction of young stars in Abell 2390. We do not fit the shape of the continuum when measuring absorption lines and fitting stellar population models, minimising the effect of dust obscuration. Our detections or non-detections of young stellar components, and the mass fractions of the young stellar components, are therefore robust. We also note that \citet{Rawle2012} find low H$\alpha$ extinction for all but the most IR-luminous BCGs, compared to similar normal star forming galaxies, implying that the gas in the BCGs is, in general, significantly less dusty.


\subsection{Host cluster X-ray properties}

There are two thresholds in the surrounding medium of the host cluster that determine whether star formation in BCGs is allowed, and at what rate: a threshold in central entropy\footnote{We use the central entropy measured by \citet{Mahdavi2013} and follow their notation of K$_{0}$ = K(20 kpc) kev cm$^{2}$.}, $K_{0}$, and a threshold in the ratio of cooling-time to free-fall time, $t_{c}/t_{ff}$.

A sharp threshold for star formation in BCGs is found at $K_{0} < 30$\ keV\ cm$^{2}$ \citep{Cavagnolo2008}. This entropy criterion divides BCGs with various observed indicators for star formation from those without it \citep{Voit2014}, and indicates that some combination of thermal conduction and suppression of thermal instability prevent condensation and star formation in systems of greater core entropy. This observed threshold also translates as a central cooling time of $t_{c, 0} < 1$ Gyr, as $t_{c, 0}$ and $K_{0}$ are closely related. Below the central entropy threshold, short cooling times enhance the likelihood of steady or intermittent gas flow from the ICM to the galaxy \citep{McCarthy2008}. As introduced in section \ref{Sec1}, this is a necessary but not sufficient condition for star formation to occur. 

Precipitation (leading to star formation) in the centres of cool-core clusters is tightly linked to the minimum ratio of the cooling time over the free-fall time, $t_{c}/t_{ff}$ \citep{Sharma2012, Gaspari2012, Voit2015, Li2015, Prasad2015}, where the cooling time, $t_{c}$, is an estimate of the time it would take for a parcel of cluster gas to radiate all of its energy at its current X-ray luminosity, $L_{X}$. The theoretical framework has clarified the conditions under which condensation of a low-entropy core can produce multiphase structure and indicates that the threshold for precipitation is $t_{c}/t_{ff} \sim$ 10 (\citealt{Singh2015}, but see \citealt{Meece2015}), and correlates with the total amount of multiphase gas \citep{Voit2014}. Thus, below the $t_{c}/t_{ff}$ threshold the thermal instabilities grow and the gas becomes multiphase gas that falls in, and fuels the AGN cycle and the star formation (\ie \ a sufficient condition for star formation to occur).

In this section, we explore how the $K_{0}$ and $t_{c}/t_{ff}$ criteria relate to the detected young stellar components, and we compare the luminosity-weighted SSP ages of the BCGs (without very young components) to the X-ray properties of the host clusters (projected distance between BCG and X-ray peak $R_{off}$, central cooling time $t_{c, 0}$, central entropy $K_{0}$, and minimum cooling to free-fall timescale ratio $t_{c}/t_{ff}$). We use the central measurements of the clusters by \citet{Mahdavi2013} from the high-resolution Chandra and XMM-Newton X-ray data for the CCCP sample.

\medskip

\begin{figure}
   \centering
   \includegraphics[scale=0.33]{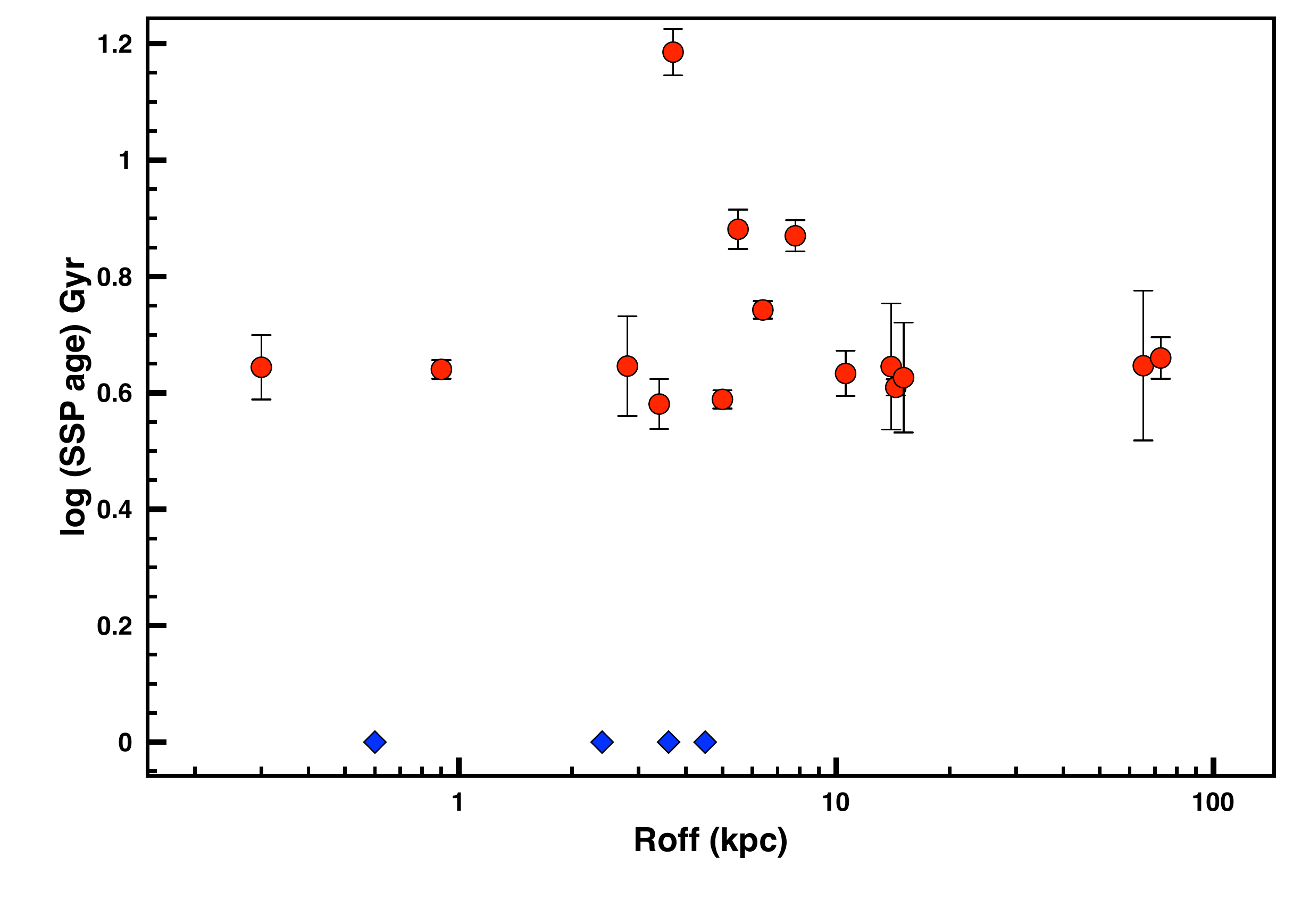}
   \caption[]{SSP ages of the inner apertures plotted against the X-ray offsets, R$_{off}$. The four galaxies with very young components (blue diamonds) are all located $<$ 5\ kpc in projection from the X-ray peak, and the X-ray offsets of these galaxies are plotted at Log (SSP age) = 0 for reference.}
   \label{Roff}
\end{figure}

In Figure \ref{Roff}, we plot the SSP luminosity-weighted ages of the inner regions against the projected distance between the BCG and the X-ray peak of the cluster, $R_{off}$ (in kpc). The four galaxies with very young components (blue diamonds) are all located $<$ 5\ kpc from the X-ray peak, and the figure emphasises this as a necessary but not sufficient condition for star formation. Whenever a significant SF is observed in the BCG, its position is within few kpc from the X-ray centre.

\begin{figure}
   \centering
   \includegraphics[scale=0.33]{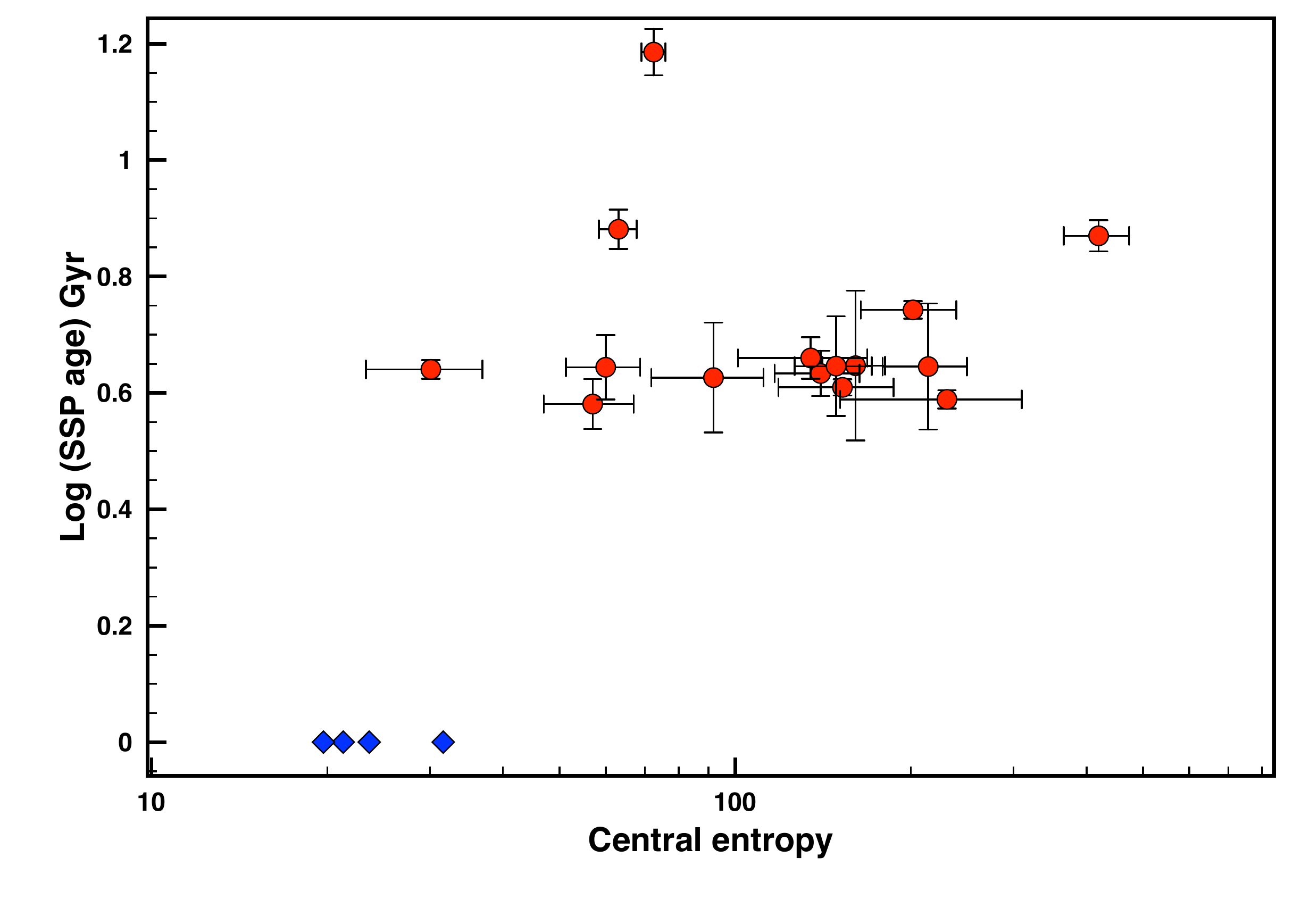}
   \caption[]{SSP ages of the inner apertures plotted against central entropy, $K_{0}$. The four galaxies with very young components all have central entropy $<$ 32\ keV\ cm$^{2}$. The central entropy of these galaxies are plotted at Log (SSP age) = 0 for reference. }
   \label{K0}
\end{figure}
\begin{figure}
   \centering
   \includegraphics[scale=0.33]{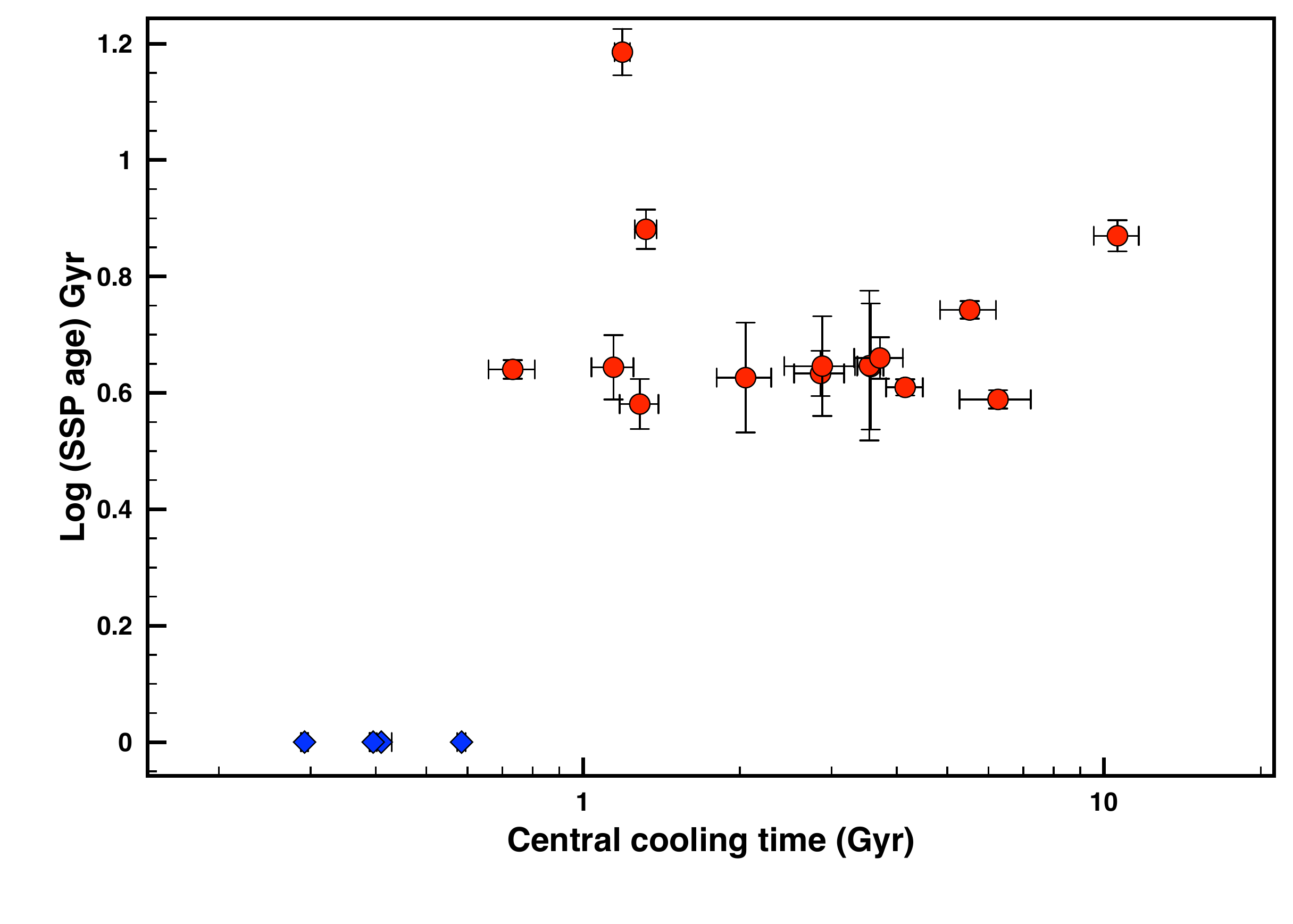}
   \caption[]{SSP ages of the inner apertures plotted against central cooling time, $t_{c, 0}$. The four galaxies with very young components all have $t_{c, 0} <$ 0.6\ Gyr, and the cooling times of these galaxies are plotted at Log (SSP age) = 0 for reference.}
   \label{tc}
\end{figure}

We plot the SSP luminosity-weighted ages of the inner apertures against the central entropy, $K_{0}$ (in keV\ cm$^{2}$), and central cooling time, $t_{c, 0}$ (in Gyr) in Figures \ref{K0}  and \ref{tc}. The four galaxies with very young components all have central entropy $<$ 32\ keV\ cm$^{2}$, and central cooling time $<$ 0.6\ Gyr. 

\begin{figure}
   \centering
   \includegraphics[scale=0.33]{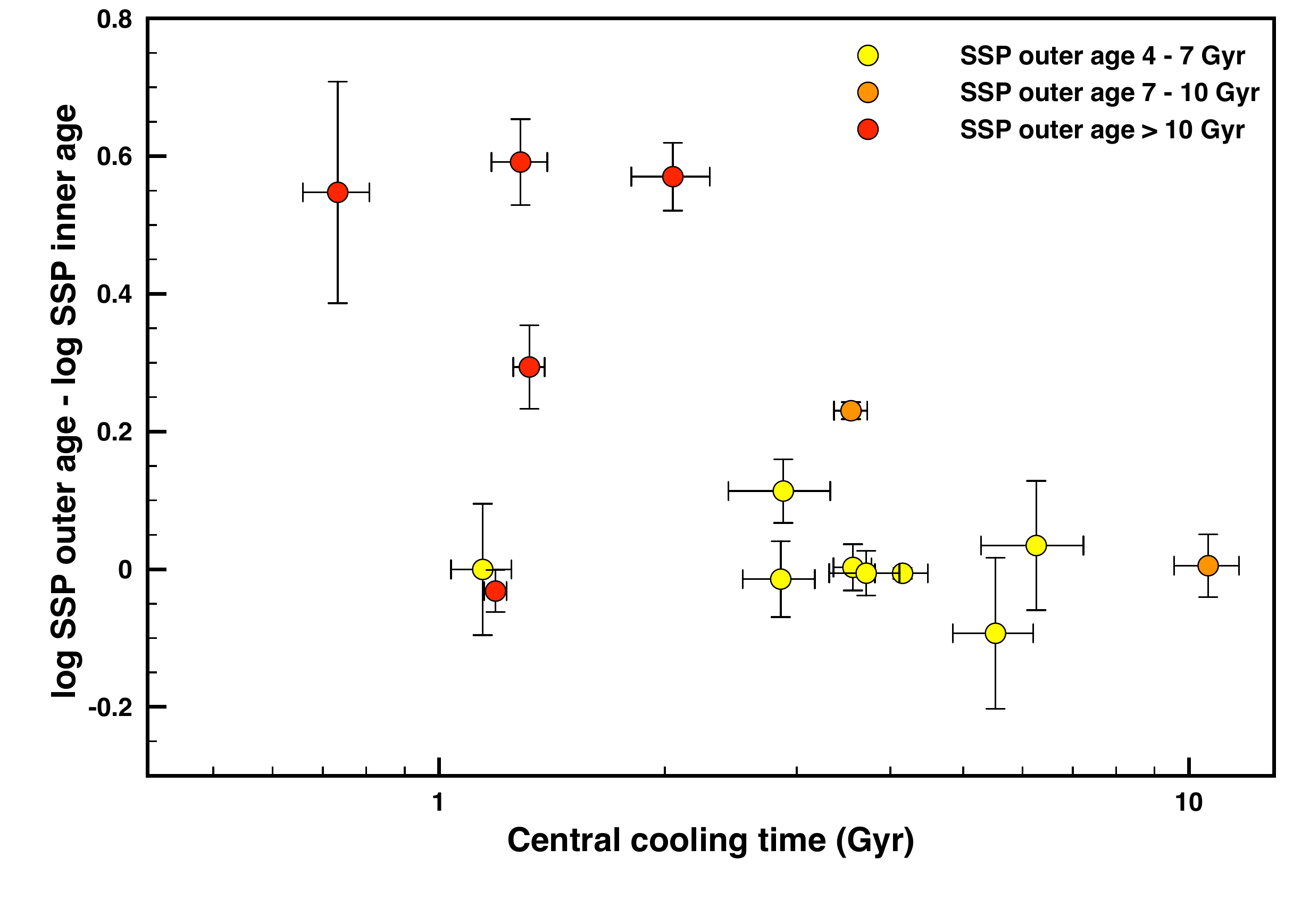}
   \caption[]{The SSP age gradients (difference between outer and inner bins) against central cooling time for all the galaxies without young components (\ie~ galaxies adequately described by a single stellar population). The symbols are colour-coded according to the SSP age of the outer bin.}
   \label{Gradtc}
\end{figure}

We also plot the SSP age gradients (log outer bin age -- log inner bin age) against cooling time for all the galaxies without a (very) young stellar component in Figure \ref{Gradtc}. There are interesting trends apparent in this plot: First, the BCGs with the oldest stars in the outer bin all reside in clusters with central cooling time $<$ 2 Gyrs while those with intermediate (4 -- 10 Gyrs) SSP ages typically live in clusters with long cooling times $>$ 2.5 Gyrs. ÊAnd second, the latter systems generally show shallow SSP age gradients, with stars in the inner bin being approximately 20-30\% younger while the BCGs residing clusters with shorter central cooling times show a much wider dispersion in the age gradients, the SSP ages of the stars in the inner being as young as a quarter of the SSP age of stars in the outer bin.

The results are intriguing, and seem to suggest that BCGs in clusters with moderate to long central cooling times are somewhat younger systems and that the shallow age gradients suggest galaxy-wide star formation perhaps associated with the last major merger that established the cluster/BCG system. On the other hand, any system-wide star formation in ÊBCGs living in the cool core clusters happened a long time ago, if at all, as indicated by the old stellar ages in the outer bin, and the dispersion in the age gradient suggests that any secondary star formation that happens is largely confined to the central regions.  In previous studies, SSP age gradients have been shown to correlate with galaxy mass and kinematics but not the local environment \citep{Kuntschner2010, McDermid2015}, as this plot seems to suggest. A thorough investigation of the intriguing relationship between age gradients and BCG dynamic properties will be presented in a follow-up study, but we have looked at the relationship between cluster mass and BCG age gradients and do not see any relationship.  The only strong trend appears to be with the central cooling time and the gas thermal instability parameter ($t_{c}/t_{ff}$) and therefore, one tantalizing possibility is that the star formation signature may possibly be related to the longer-term heating-cooling cycles in the cluster cores. Regardless, Figure \ref{Gradtc} shows that a spatially-resolved SSP analysis offers the possibility to resolve star formation histories in BCGs despite the degeneracy that causes composite stellar population models to be unreliable for BCGs without star formation in the last Gyr.

\begin{figure}
   \centering
   \includegraphics[scale=0.33]{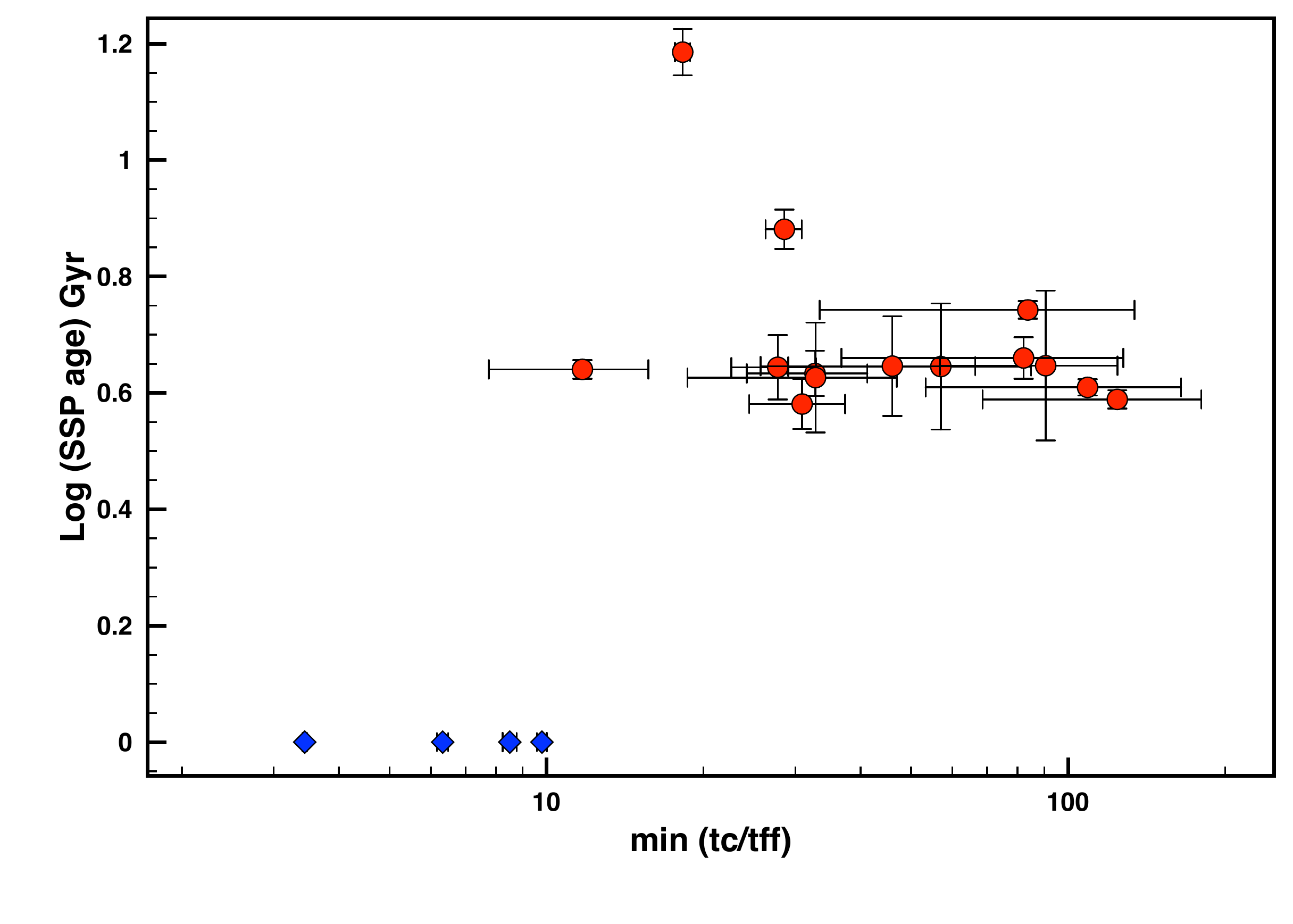}
   \caption[]{SSP ages of the inner apertures plotted against minimum ratio $t_{c}/t_{ff}$. The minimum $t_{c}/t_{ff}$ of the four BCGs with young components are shown at Log (SSP age) = 0 for reference.}
   \label{tff}
\end{figure}

We calculate the $t_{c}/t_{ff}$ ratio by using the redshift of the BCGs, the $M_{500}$ lensing mass of the clusters \citep{Hoekstra2015}, the $t_{c, 0}$ (as described above) and the X-ray derived cluster concentration value \citep{Mahdavi2013}, assuming an NFW profile \citep{Navarro1996}. The $M_{500}$ mass is the mass enclosed within a radius where the mean density is 500 times the critical density\footnote{All calculations are done assuming H$_{0}$ = 70\ km\ s$^{-1}$ Mpc$^{-1}$.}. One caveat of the NFW assumption (that does not include the potential of the central BCG) is that it can lead to smaller values of $t_{c}/t_{ff}$. We propagate the measurement errors, and we find a good agreement with this derivation of the $t_{c}/t_{ff}$ ratio compared to the calculations in \citet{Voit2014} for the CCCP clusters in common with the ACCEPT cluster catalogue \citep{Cavagnolo2009}. We find one value (for the BCG in Abell 1763) that is extremely large due to an unusually small concentration value and exclude it from further analysis. In Figure \ref{tff}, we compare the SSP ages for the inner apertures with the $t_{c}/t_{ff}$ ratio, where we also plot the the $t_{c}/t_{ff}$ ratios of the four BCGs with prominent young components at Log (SSP age) = 0 for comparison. Our results show that the four BCGs with young components are all located below the threshold for precipitation of $t_{c}/t_{ff} \sim$ 10. 

It is still possible that gas is present in BCGs with $t_{c}/t_{ff} >$ 10. This is because the cold precipitation and the ensuing AGN feedback, which can intermittently raise the central cooling time, proceeds in cycles. Within this cycle, there can be a significant delay\footnote{This is model dependent. The delay in the simulations of \citet{Prasad2015} is shorter than those in \citet{Li2015} because of smaller jet efficiencies.} of up to 2 Gyr between when the ICM cooling rate is suppressed (increase in $t_{c}/t_{ff}$) due to AGN heating and when SFR starts to decline at the end of the cycle due to the consumption of the cold gas. The simulations of \citet{Li2015} show that $t_{c}/t_{ff}$ can fluctuate between $\sim$ 5 and $\sim$ 20 while star formation is happening. Therefore, the relation between SFR and gas cooling rate (or $t_{c}/t_{ff}$) at a particular moment in time is not linear, but the two quantities are related in that star formation generally occurs only when $t_{c}/t_{ff}$ is close to the critical value for precipitation \citep{Voit2015, Li2015}.

We would like to draw attention to the results of MS 0440+02, where $R_{off} \sim 0.9$ kpc, and $S_0 \sim 30\;\rmn{keV\, cm}^2$, and $t_{c}/t_{ff}$ barely above 10, but where no prominent young stellar component is detected. This non-detection of a young component is unconventional given the combination of the three properties. This is one of the four BCGs in this sample (Abell 611, 963, 2537, MS 0440+02) with a large age gradient between the luminosity-weighted SSP age of the inner and outer apertures. The outer apertures indicate that the majority of the stars are $\sim 14$ Gyr old, and all four host clusters have cooling times below $\sim 2$ Gyr (see Figure \ref{Gradtc}). It is thus very likely that these galaxies (and in particular MS 0440+02) have had active evolutionary paths and possibly bursts of star formation over cosmic time, generating one or more younger (but older than 2 Gyr) stellar population components superposed on an old population. It is worth mentioning that this system was found to have a red core in \citet{Bildfell2008} but with very high B-band luminosity. \citet{Chapman2002} had previously found a significant amount of red dust in the galaxy and the high resolution imaging in \citet{Bildfell2008} showed a double nucleus feature that could be due to the presence of a dust lane. \citet{Bildfell2008} speculated that the system may be hosting dust-shrouded star formation, although the study by \citet{Hoffer2012} did not detect dust-obscured star formation. From the X-ray properties and theoretical framework, you would expect to see star formation. Given our analysis, dust is unlikely to obscure a young stellar component, but the absence of a young stellar component can be due to the potential for hysteresis in the precipitation cycle (\citealt{Voit2015, Li2015}). \citet{Voit2014} discuss the absence of star formation in low entropy systems, as conduction can potentially prevent gas from condensing in the cores.

As mentioned in section 2, the clusters that we did not observe in the allocated time are Abell 2204, Abell 2218, Abell 2219, Abell 521, Abell 697. These are all listed as red cored galaxies in \citep{Bildfell2008}, except for Abell 2204 which is a known blue core BCG. According to this, and their environment characteristics as listed in Table \ref{bcgs}, these galaxies fit the trends described above and we conclude that these systems would sit alongside the other systems with similar characteristics.

\section{Summary and conclusions} 

We investigate the stellar population properties in BCGs located in the very centres of a subset of the CCCP sample of X-ray luminous clusters at redshift 0.15 $< z <$ 0.3. Specifically, we observe and analyse the absorption line spectra in the blue optical wavelength range. The latter confers the advantage of us being able to robustly identify very young $<$ 1\ Gyr populations in some of these BCGs. We also find a large variety in stellar population properties, and most likely SFHs, in this sample of BCGs that indicates a diverse set of complex, non-passive, evolutionary paths.

We detect prominent young ($\sim$ 200 Myr) stellar populations in 4 of the 19 galaxies. We constrain the mass contribution of these young components to the total stellar mass to be typically between 1\% to 3\%. The BCG in Abell 1835 is, however, unusual in that it shows remarkable A-type stellar features indicating a relatively large population of young stars (7\% of the total stellar mass), which is extremely unusual even amongst star forming BCGs.  We also show that a spatially-resolved SSP analysis of stellar age gradients within individual BCGs offers the possibility of resolving star formation histories in BCGs without star formation in the last Gyr despite the degeneracy that causes composite stellar population analyses to become unreliable in such systems.

We also explore the relationship between the star formation history and the local environment of the BCGs. We find that the four BCGs with strong evidence for recent star formation (and only these four galaxies) are found within a projected distance of 5 kpc of their host cluster's X-ray peak, \textit{and} the diffuse, X-ray gas surrounding the BCG exhibits a ratio of the radiative cooling-to-free-fall time of $t_{c}/t_{ff} < $ 10. These galaxies also have some of the lowest central entropy, all hosted by clusters with $S_0 < 32\;\rmn{keV\, cm}^2$. The stellar population results are consistent with a UV photometric analysis, in showing a large variety of SFHs \citep{Donahue2015}, as well as the age and mass contribution of the young components \citep{Pipino2009} in clusters with low entropy, thermally unstable cores. 

Our results are consistent with the predictions of the intermittent precipitation-driven star formation and AGN feedback model, in which the radiatively cooling diffuse gas is subject to local thermal instabilities once the instability parameter $t_{c}/t_{ff}$ falls below $\sim$ 10, leading to the condensation and precipitation of cold gas. In this model, radiative cooling of the thermally unstable gas causes the condensation and precipitation of cool clouds out of the hot medium surrounding the BCG, which then rain down onto the BCG and fuel star formation as well as the central supermassive black hole. The resulting AGN feedback heats the diffuse gas, temporarily suppressing the further formation of the thermal instabilities. This, in turn, eventually leads to a shutdown of AGN feedback and star formation, setting the stage for the next cycle.

Consider a population of BCGs where all the conditions ($R_{off}$, $K_{0}$ and $t_{c}/t_{ff}$) that will make the BCGs susceptible to the cycles of star formation are satisfied. In the event of continuous, or recurring bursts of star formation repeating on a timescale of 200 Myr or shorter, the youngest ($\sim$ 100 -- 200 Myr), most luminous stellar component is the most dominant young component. On the other hand, if the burst recurrence timescale is of the order of $\sim$ 1 Gyr, then we find that the young component will only be detected for 20\% of the time, and a component between 200 -- 1000 Myr will be observed 80 \% of the time. Our sample of BCGs, where all four BCGs that reside in host clusters that satisfy all the criteria from star formation episodes show a very young stellar component, suggest a timescale for star formation to restart on the order of $\sim$ 200 Myrs, or less.

\section*{Acknowledgments}

This research was enabled in part by support provided by (a) the National Science Foundation under Grant No. 1066293 to the Aspen Center for Physics, (b) the National Research Foundation of South Africa to SIL, (c) NSERC (Canada) Discovery Grant to AB, (d) Chandra Archive grant AR3-14013X, HST grant STSI HST-GO-12065.07-A and NASA ADAP grant NASA NNX13AI41G to MD.

We thank the anonymous referee for the constructive comments on the manuscript. SIL thanks Megan Donahue and Mark Voit for hosting her visit to Michigan State University and to the American Physical Society partially funding the visit.  AB, MD and MV would like to acknowledge the simulating collaborative atmosphere of the Aspen Center for Physics, where some of the work reported here was carried out with the support of  National Science Foundation grant PHY-1066293.   

Observational data used in this article obtained on Gemini North and South telescope under proposal IDs: GS--2007B--Q--36; GN--2008A--Q--103; GS--2008B--Q--21; GN--2008B--Q--5; GS--2008B--Q--4 (PI: C. Bildfell). This research used the facilities of the Canadian Astronomy Data Centre operated by the National Research Council of Canada with support from the Canadian Space Agency.   

Any opinion, finding and conclusion or recommendation expressed in this material is that of the author(s) and the NRF does not accept any liability in this regard.

\bibliography{References}

\bsp

\label{lastpage}

\end{document}